\documentclass{ws-ijmpa}
\usepackage[super,compress]{cite}
\usepackage{graphicx}

\catchline{}{}{}{}{}
\usepackage{amsmath,amssymb} 
\usepackage{bm}
\usepackage{bbm}
\usepackage{cite}
\usepackage{color} 
\usepackage{graphicx}
\usepackage{tabularx}

\usepackage{epstopdf}
\newenvironment{noteblue}{\color{blue}} {\ignorespacesafterend}

\newcommand{\Sec}[1]{Sec.~\ref{#1}}  

\renewcommand{\figurename}{FIG.}

\newcommand{\fig}[1]{Fig.~\ref{#1}}

\renewcommand{\tablename}{TAB.} 

\newcommand{\tab}[1]{Tab.~\ref{#1}}

\newcommand{\eq}[1]{Eq.~(\ref{#1})}

\begin{document}
\markboth{Hong-Bo Jin, Yue-Liang Wu and Yu-Feng Zhou}{Implications of the first AMS-02 data for dark matter}
\title{
Implications of the first AMS-02 antiproton data for dark matter
}
\author{
Hong-Bo Jin$^{a,b}$, 
Yue-Liang Wu$^{a,c,d}$, \ 
and  Yu-Feng Zhou$^{a,c}$
}

\address{
\it $^{a)}$ State Key Laboratory of Theoretical Physics,
\it $^{b)}$ National   \\ 
\it Astronomical  Observatories, 
\it Chinese Academy of Sciences, \\
\it $^{c)}$ Kavli Institute for Theoretical Physics China,
\it Institute   of  \\
\it Theoretical Physics Chinese Academy of Sciences,\\
\it $^{d)}$University of Chinese Academy of Sciences,\\ 
\it Beijing, 100190, P.R. China
}
%

\maketitle
%
%
\begin{abstract}
The implications of the first AMS-02 $\bar p/p$ data for
the propagation of cosmic rays and the properties of dark matter (DM) are  discussed.
Using various diffusive re-acceleration (DR) propagation models, 
one can derive very conservative upper limits on 
the DM annihilation cross sections.
The limits turned out to be compatible with that from 
the Ferm-LAT gamma-ray data on 
the dwarf spheroidal satellite galaxies.
The flattening of the $\bar p/p$ spectrum above  $\sim 100$~GeV in the current data still 
leaves some room for TeV scale DM particles.
More antiproton data at high kinetic energies are needed to constrain the properties of the 
DM particles.
\end{abstract}


	
\newpage

%
%
%

Cosmic-ray antiparticles, such as positrons and antiprotons play important roles
in the indirect search for dark matter (DM) in the Galactic halo.
The Alpha Magnetic Spectrometer (AMS-02)   is measuring 
such cosmic-ray charged particles with unprecedented accuracies.
So far the  anomalous rise in the positron fraction previous reported by 
PAMELA~\cite{
Adriani:2008zr,
Adriani:2010ib
}
and Fermi-LAT~\cite{
FermiLAT:2011ab
}
has been confirmed by AMS-02 with higher accuracy and
extended to higher energies
\cite{Accardo:2014lma}, 
which has triggered  extensive theoretical studies on 
possible explanations including halo DM annihilation or decay
 (for recent global analyses on AMS-02 data, see e.g. Refs 
 \cite{
Kopp:2013eka,
DeSimone:2013fia,
Cholis:2013psa,%
Feng:2013zca, 
Jin:2013nta,Liu:2013vha,%
Bergstrom:2013jra,
Ibarra:2013zia,
DiMauro:2014iia,
%
Lin:2014vja,Ibe:2014qya%
} ).
Antiprotons are highly expected from DM annihilation in many DM models,
which is  unlikely to be generated from the nearby pulsars.
%

Recently, the AMS-02 collaboration has released 
the first preliminary result of  the cosmic-ray antiproton to proton flux ratio $\bar p /p$
\cite{Ting:AMS}.
The measured  kinetic energies of the antiprotons have been extended to  
$\sim 450$ GeV. 
%
Although 
the spectrum of $\bar p /p$ at high energies above 100 GeV tend to be relatively flat,
within uncertainties 
the AMS-02 data are consistent with 
the  background of secondary antiprotons,
which can be used to set stringent upper limits on 
the  dark matter (DM) annihilation cross sections,
especially for high mass DM particles.
%
The constraints on the DM properties from antiprotons have been
investigated previously before  AMS-02
( see e.g.
\cite{Hooper:2014ysa,
%
Kappl:2014hha,
%
Fornengo:2013xda,
%
Cirelli:2013hv,
%
Jin:2012jn
} ).
In this talk, 
we briefly summarise  our work on 
the implications  of the new AMS-02 $\bar p/p$ data for constraining the 
annihilation cross sections of the DM particles in 
various propagation models and DM profiles. 
The details of the analysis can be found in Ref.%
~\cite{Jin:2015sqa}
%
%


%
In the diffusion models of cosmic-ray propagation, 
the Galactic halo within which  the diffusion processes occur is parametrized by 
a cylinder with radius $R_{h} = 20-30$ kpc and half-height $Z_{h}=1-20$ kpc.   
%
%
The diffusion  equation for the cosmic-ray charged particles reads
%
\begin{align}\label{eq:propagation}
  \frac{\partial \psi}{\partial t} =&
  \nabla (D_{xx}\nabla \psi -\boldsymbol{V}_{c} \psi)
  +\frac{\partial}{\partial p}p^{2} D_{pp}\frac{\partial}{\partial p} \frac{1}{p^{2}}\psi
  -\frac{\partial}{\partial p} \left[ \dot{p} \psi -\frac{p}{3}(\nabla\cdot \boldsymbol{V}_{c})\psi \right]
  \nonumber \\
  & -\frac{1}{\tau_{f}}\psi
  -\frac{1}{\tau_{r}}\psi
  +q(\boldsymbol{r},p)  ,
\end{align}
where $\psi(\boldsymbol{r},p,t)$ is  
the number density per unit of total particle momentum.
%
For steady-state diffusion, it is assumed that  $\partial  \psi/\partial t=0$.
The number densities of cosmic-ray particles are assumed to be vanishing at 
the boundary of the halo.
%
%
%
The energy dependent spatial diffusion coefficient $D_{xx}$  is parametrized as
%
$D_{xx}=\beta D_{0} \left(\rho/\rho_{0}\right)^{\delta}$ ,
where $\rho$ is the rigidity of the cosmic-ray particle.
The  power spectral index $\delta$ can have  different values 
$\delta=\delta_{1(2)}$  
for  $\rho$  below (above) a reference rigidity $\rho_{0}$.  
%
$D_{0}$ is a normalization constant.
%
%
The convection term in the diffusion equation is related to 
the drift of cosmic-ray particles from 
the Galactic disc due to the Galactic wind.  
%
%
%
The  diffusion in momentum space is described by 
the reacceleration parameter $D_{pp}$ 
which is related to 
the Alfv$\grave{\mbox{e}}$n speed $V_{a}$ of disturbances in the hydrodynamical plasma 
\cite{Ginzburg:1990sk}.
%
The momentum loss rate is denoted  by $\dot{p}$, 
%
and $\tau_{f}(\tau_{r})$ is 
the time scale for fragmentation (radioactive decay) of 
the cosmic-ray nuclei.
 %
%

The spectrum of a primary source term for a cosmic-ray nucleus  $A$  is 
assumed to have a broken power low behaviour 
%
$dq_{A}(p)/dp \propto
\left( \rho/\rho_{As}\right)^{\gamma_{A}}$
with $\gamma_{A}=\gamma_{A1}(\gamma_{A2})$ for 
the nucleus rigidity $\rho$ below (above) a reference rigidity $\rho_{As}$.
%
%
%
The spatial distribution of the primary sources is assumed to 
follow that of the pulsars and is taken  from
Ref.~\cite{Strong:1998pw}.
 %
%
%
The background antiproton is assumed to only have the secondary origin,
namely,
they are created dominantly from 
inelastic $pp$- and $p$A-collisions with the interstellar  gas.
The corresponding source term  reads
\begin{align}
q_{\text{sec}}(p)=
\beta c n_{i} \sum_{i=\text{H,He}}
\int dp'   \frac{\sigma_{i}(p,p')}{dp'} n_{p}(p') ,
\end{align}
where 
$n_{i}$ is the number density of the interstellar hydrogen (helium),
$n_{p}$ is the number density of primary cosmic-ray proton per total momentum, 
and $d\sigma_{i}(p,p')/dp'$ is the differential cross section
for  $p+\text{H(He)}\to \bar p + X$. 
%
%
In calculating the antiprotons, 
inelastic scattering to produce ``tertiary'' antiprotons should be 
taken into account.
%
%
%
%
%
The primary source  from 
the annihilation of Majorana DM particles has the  following form
\begin{align}\label{eq:ann-source}
q_{\text{\tiny DM}}(\boldsymbol{r},p)=\frac{\rho(\boldsymbol{r})^2}{2 m_{\chi}^2}\langle \sigma v \rangle 
\sum_X \eta_X \frac{dN^{(X)}}{dp} ,
\end{align}
where $\langle \sigma v \rangle$ is 
the velocity-averaged DM annihilation cross section multiplied by DM relative velocity.
%
$\rho(\boldsymbol{r})$ is the DM energy density distribution function,
and
$dN^{(X)}/dp$ is the injection energy spectrum  of  antiprotons
from DM annihilating into SM final states through 
possible  intermediate states $X$ with  
$\eta_X$ the corresponding branching fractions. 
The   interstellar flux of the cosmic-ray particle is related to its density function as 
%
$\Phi= v\psi(\boldsymbol{r},p)/(4\pi)$.
%
At the top of the atmosphere (TOA) of the Earth, 
the fluxes of cosmic-rays  are affected  by solar winds 
and the helioshperic magnetic field. 
This effect is taken into account using 
the force-field approximation
which involves the Fisk potential $\phi$%
~\cite{Gleeson:1968zza}.
%
%
%
We shall take  $\phi=550$~MV in numerical analysis.

\begin{figure}[tb]
\begin{center}
\includegraphics[width=0.65\textwidth]{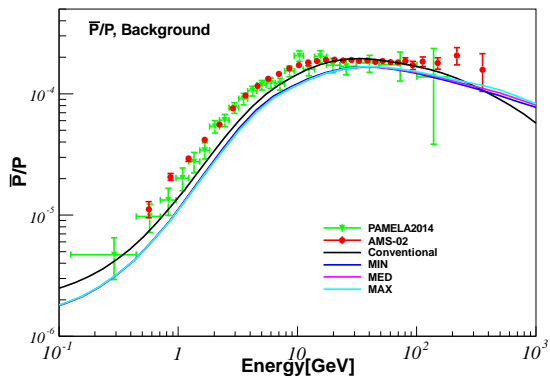}
\caption{
Predictions for the $\bar p /p$ ratio from the four propagation models.
%
The data  from AMS-02~\cite{Ting:AMS}
and PAMELA~\cite{Adriani:2014pza} are shown.
See text for detailed discription.
}
\label{fig:pbar}
\end{center}
\end{figure}

We  solve the diffusion equation of \eq{eq:propagation} 
using the publicly available  code  GALPROP v54
\cite{astro-ph/9807150,astro-ph/0106567,astro-ph/0101068,astro-ph/0210480,astro-ph/0510335}
which utilizes  realistic astronomical information on 
the distribution of interstellar gas and other data as input, 
and considers various kinds of observables
in a self-consistent way. 
%
%
%
%
%
%
We start with
the so-called ``conventional''  diffusive re-acceleration (DR) model 
\cite{astro-ph/0101068,astro-ph/0510335}
which is commonly adopted by the current experimental collaborations
as a benchmark model for the  astrophysical backgrounds.
%
%
%
%
It is useful to consider this model as a reference model to 
understand how the DM properties could be  constrained by 
the AMS-02 data.
Then  we consider 
three representative propagation models  selected from 
a large sample of models obtained from 
a global Bayesian MCMC fit to  
the preliminary AMS-02 proton and B/C data using 
the GALPROP code
\cite{Jin:2014ica}.
They are selected  to represent the typically
minimal (MIN), median (MED) and 
maximal (MAX) antiproton fluxes within $95\%$ CL,
corresponding to the region enveloping
$95\%$ of the MCMC samples with highest likelihoods  in 
a six-dimensional parameter space. 
%
%
%
%
Note that the GALPROP based ``MIN'', ``MED'' and ``MAX'' models used in this work are
different from and  complementary to 
that given in Ref.%
~\cite{Donato:2003xg}.
The predictions for the background of the $\bar p/p$ flux ratio in these models are 
shown in \fig{fig:pbar}.
The ``MIN'', ``MED'' and ``MAX'' models are highly degenerate in the background 
$\bar p/p$ ratio.
Compared with these models, 
the ``conventional'' model predicts more low energy antiprotons but 
at high energies above $\sim 500$~GeV, the predicted antiprotons are much less. 
%
In all the four DR propagation models, below $\sim10$~GeV 
the GALPROP based calculations  underpredict
the  $\bar p/p$ flux ratio by $\sim 40\%$, which is a known issue.
%
%
The agreement with the low energy $\bar p$ data can be improved by
introducing breaks in diffusion coefficients
\cite{Moskalenko:2001ya}, 
%
``fresh'' nuclei component
\cite{Moskalenko:2002yx} 
%
or a DM contribution
\cite{Hooper:2014ysa}. 
%
%
The predictions for low energy $\bar p/p$ ratio can also be easily modified  by 
introducing an independent Fisk potential $\phi$ for $\bar p$ and  
an energy-dependent overall normalization factor as 
discussed  in Ref.\cite{Giesen:2015ufa}.
 %
We instead use these DR models to 
derived  very conservative upper limits on 
the annihilation cross sections of light DM particles.
Note, however, that in the DR propagation models,
the background predictions agree with the AMS-02 data well
at higher energies $\sim 10-100$ GeV,
which  can be turned into stringent constraints on 
the nature of heavy DM particles.
We consider three reference DM annihilation channels 
$\bar\chi\chi \to XX$ 
where $XX=q \bar q$, $b \bar b$ and $W^{+}W^{-}$.
The energy spectra of these channels are similar at high energies.
The main difference is in the average number of total antiprotons $N_{X}$
per DM annihilation of each channel.
%
%
%
%
The injection spectra $dN^{(X)}/dp$ from DM annihilation are calculated using 
the numerical package PYTHIA~v8.175
\cite{
Sjostrand:2007gs
}.

%
\begin{figure}[tb]
\begin{center}
\includegraphics[width=0.45\textwidth]{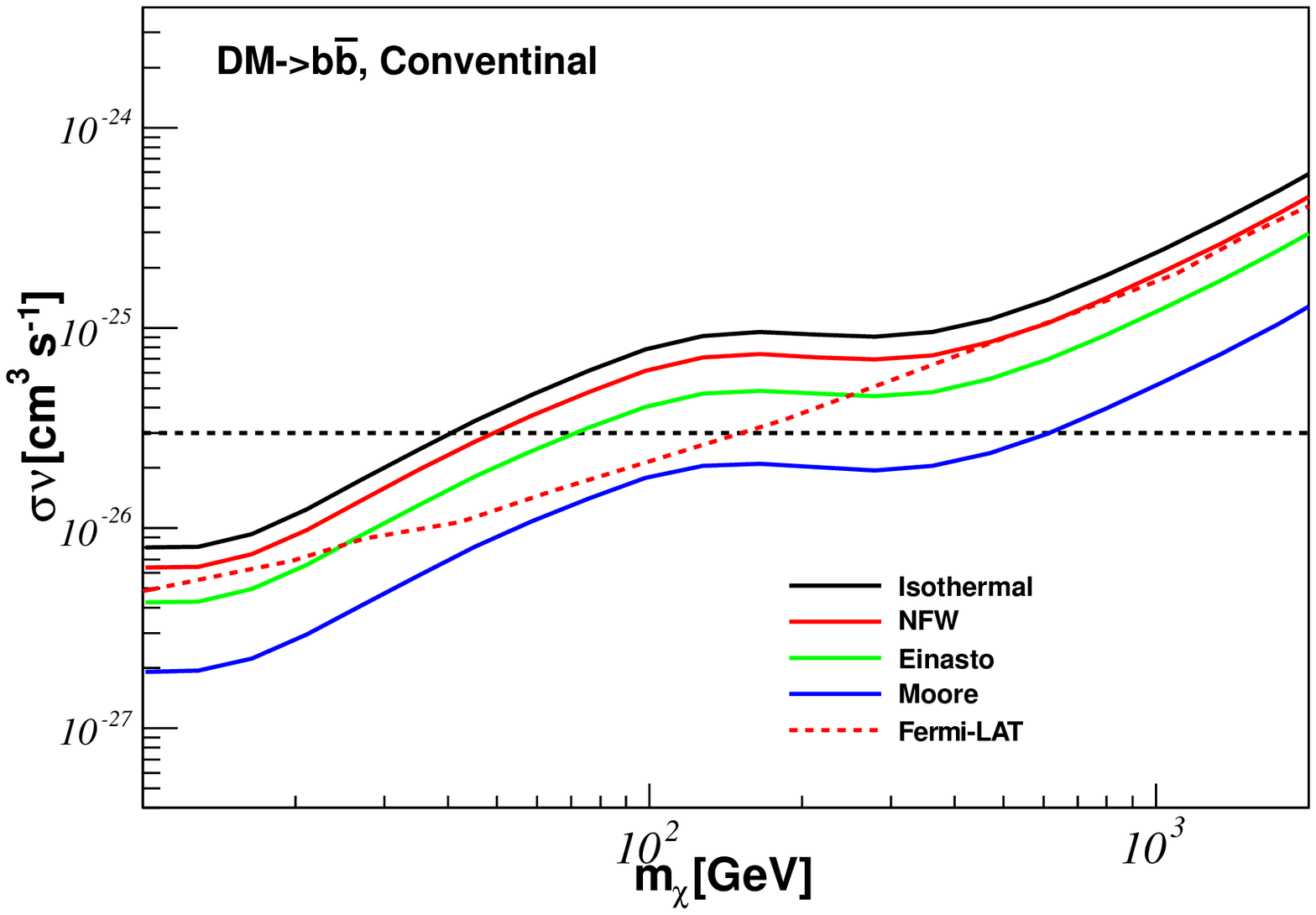}
\includegraphics[width=0.45\textwidth]{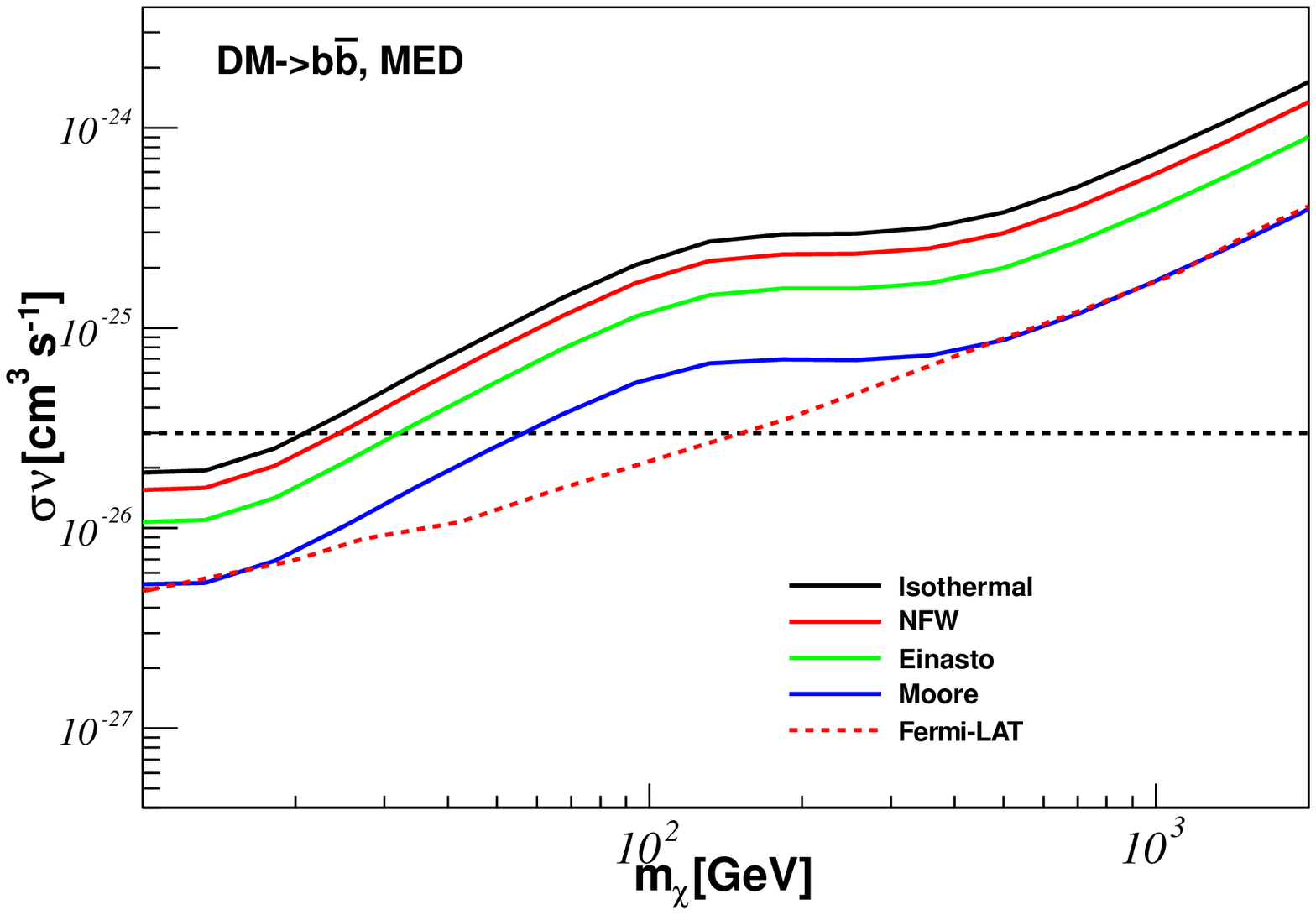}
\\
\includegraphics[width=0.45\textwidth]{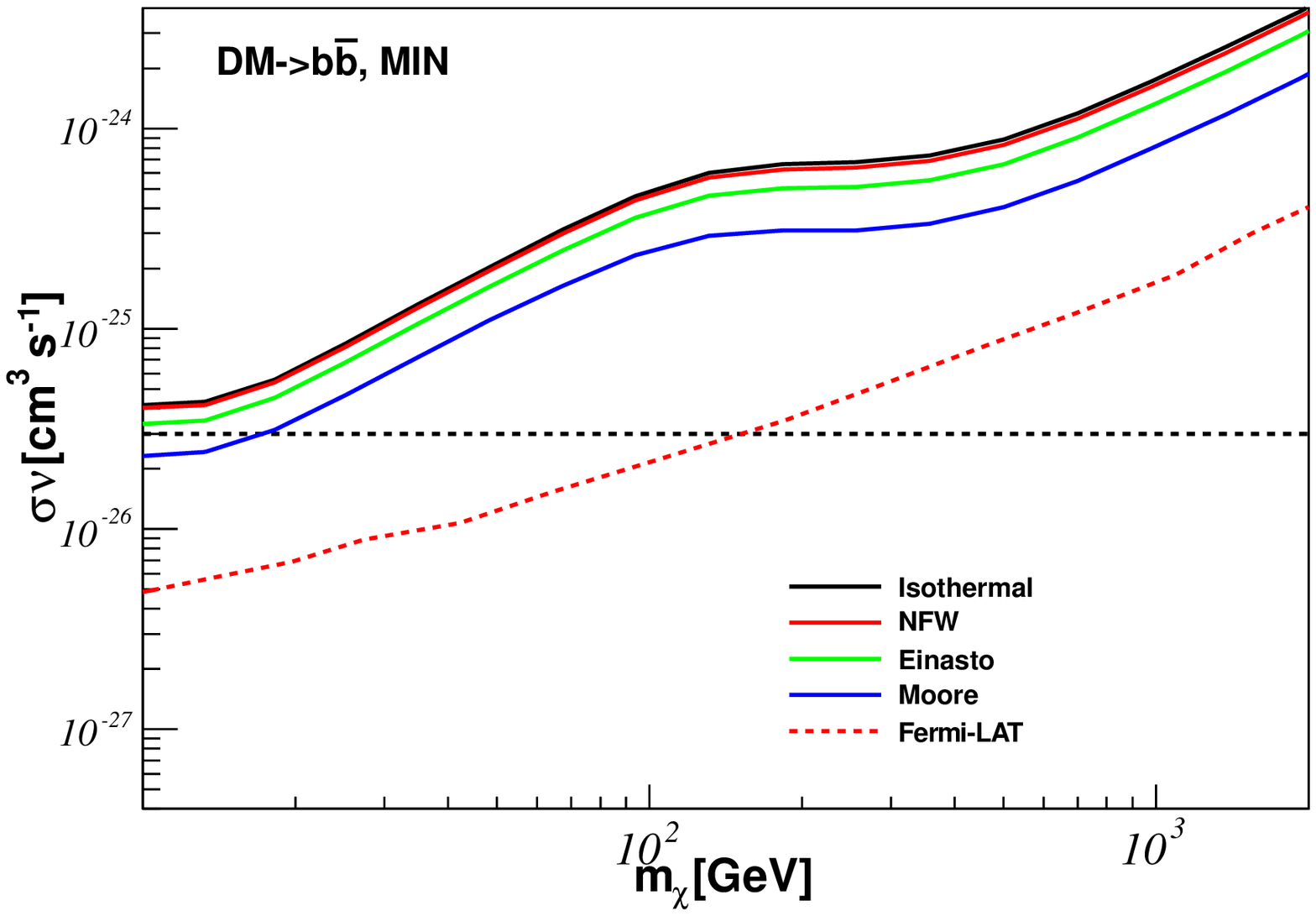}
\includegraphics[width=0.45\textwidth]{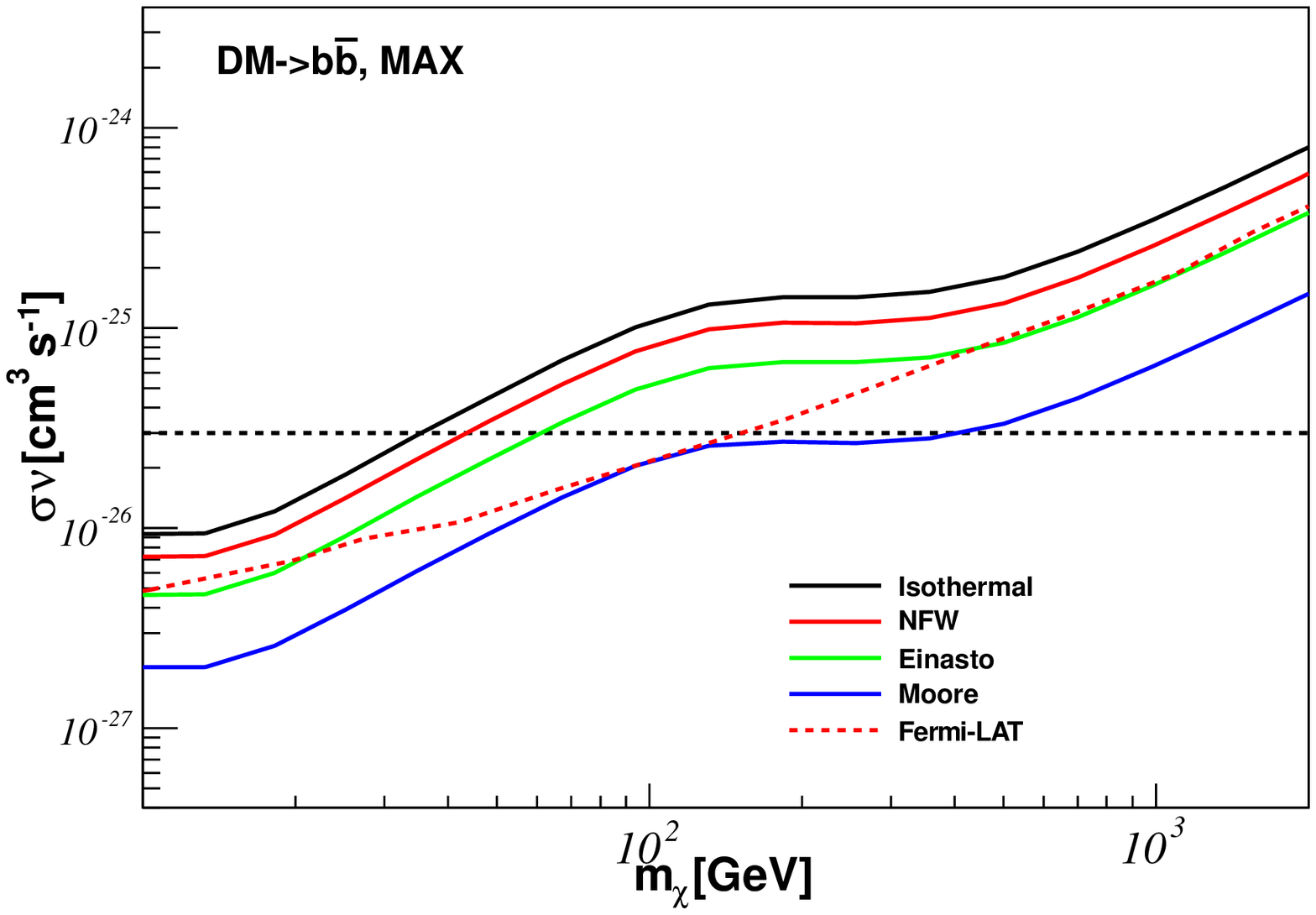}
\caption{
Upper limits on 
the cross sections for 
DM particle annihilation into $b\bar b$ final states from
the AMS-02 $\bar p/p$ data in 
the ``conventional'' (upper left),
``MED'' (upper right),
``MIN'' (lower left) and
``MAX'' (lower right)  propagation models.
Four DM profiles 
NFW~\cite{Navarro:1996gj}, 
Isothermal~\cite{Bergstrom:1997fj} , 
Einasto 
\cite{Einasto:2009zd
}
and Moore 
\cite{Moore:1999nt, 
%
Diemand:2004wh} 
are considered.
The upper limits  from 
the Fermi-LAT 6-year gamma-ray data of 
the dwarf spheroidal satellite galaxies of the Milky Way
are also shown
\cite{Ackermann:2015zua}.
The horizontal line indicates the typical thermal annihilation cross section
$\langle \sigma v \rangle=3\times 10^{-26}\text{cm}^{3}\text{s}^{-1}$.
}
\label{fig:bb}
\end{center}
\end{figure}
%

%
We shall first derive upper limits on  DM annihilation cross section 
as a function of DM particle mass,
using the frequentist $\chi^{2}$-analyses.
%
All of the 30 data points of the AMS-02 $\bar p/p$ data are included in calculating 
the limits.
%
\begin{figure}[tb]
\begin{center}
\includegraphics[width=0.45\textwidth]{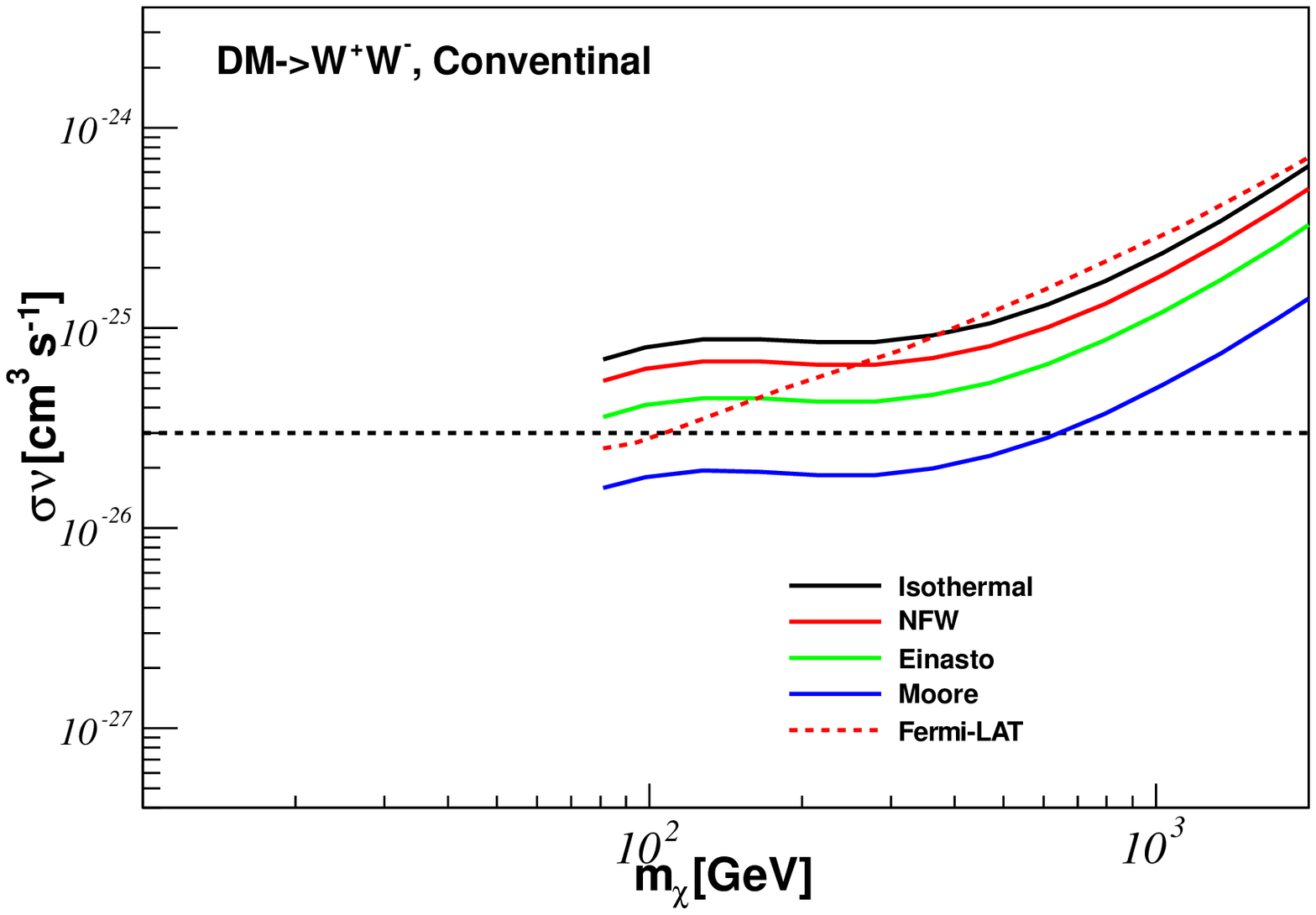}
\includegraphics[width=0.45\textwidth]{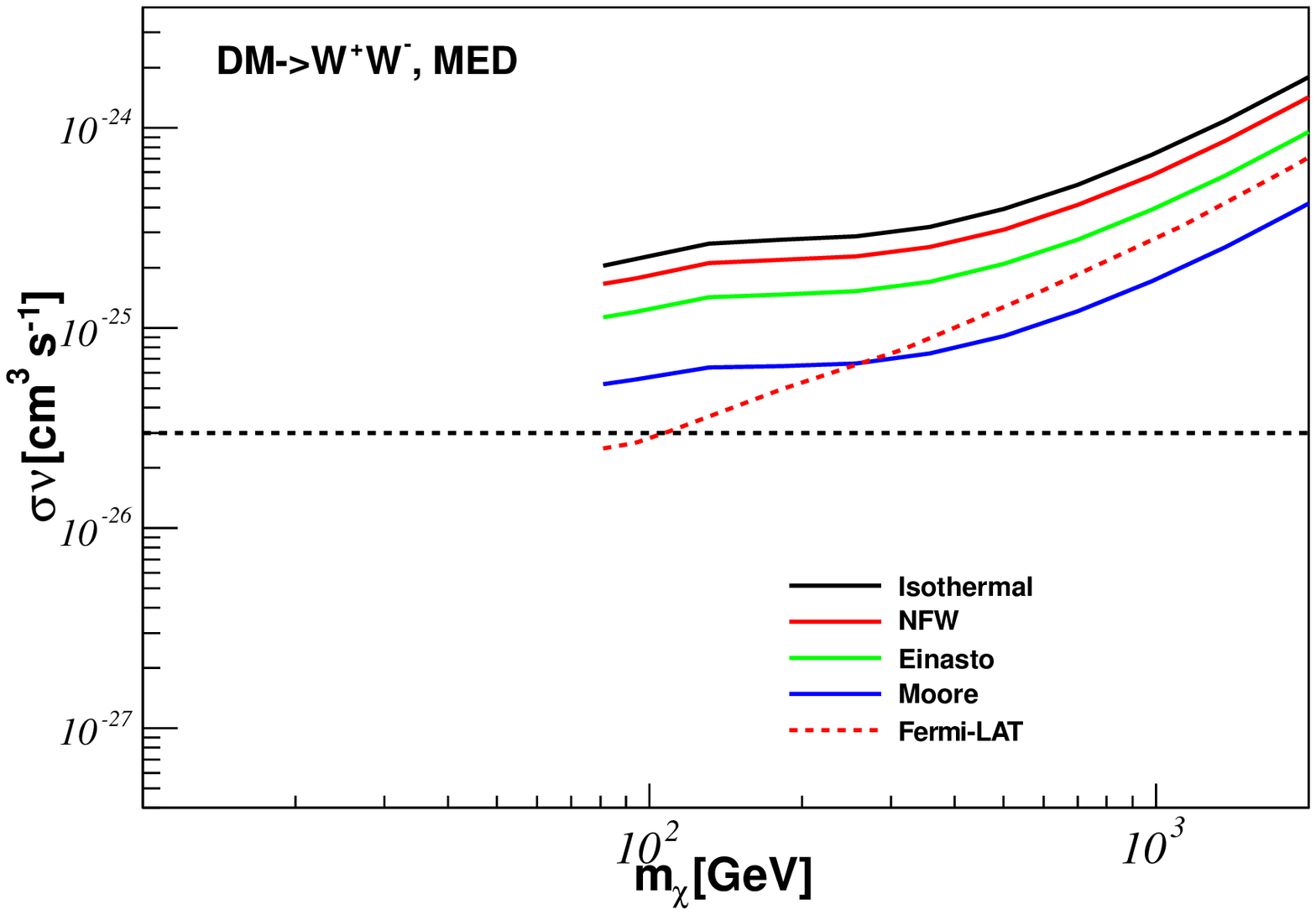}
\\
\includegraphics[width=0.45\textwidth]{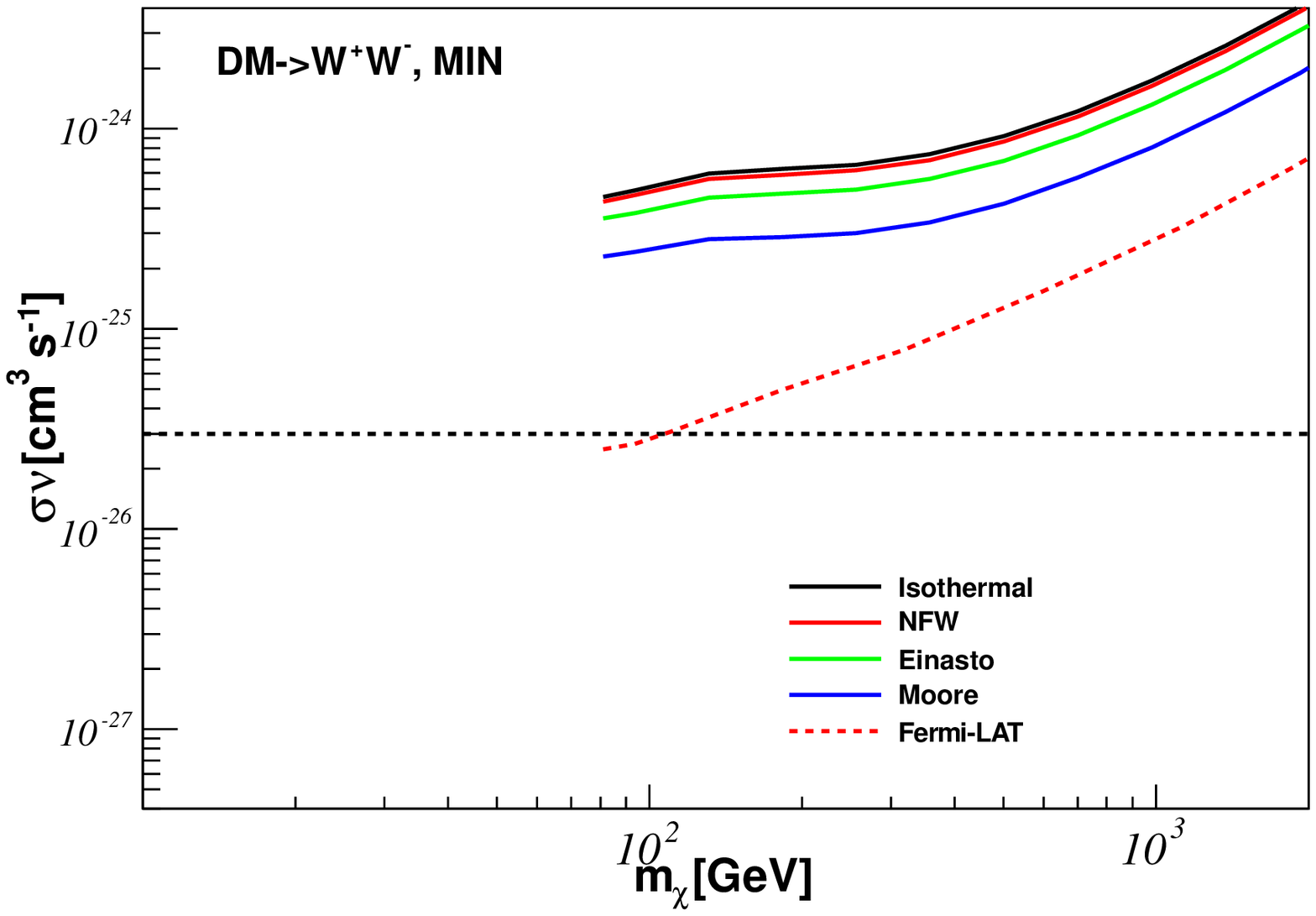}
\includegraphics[width=0.45\textwidth]{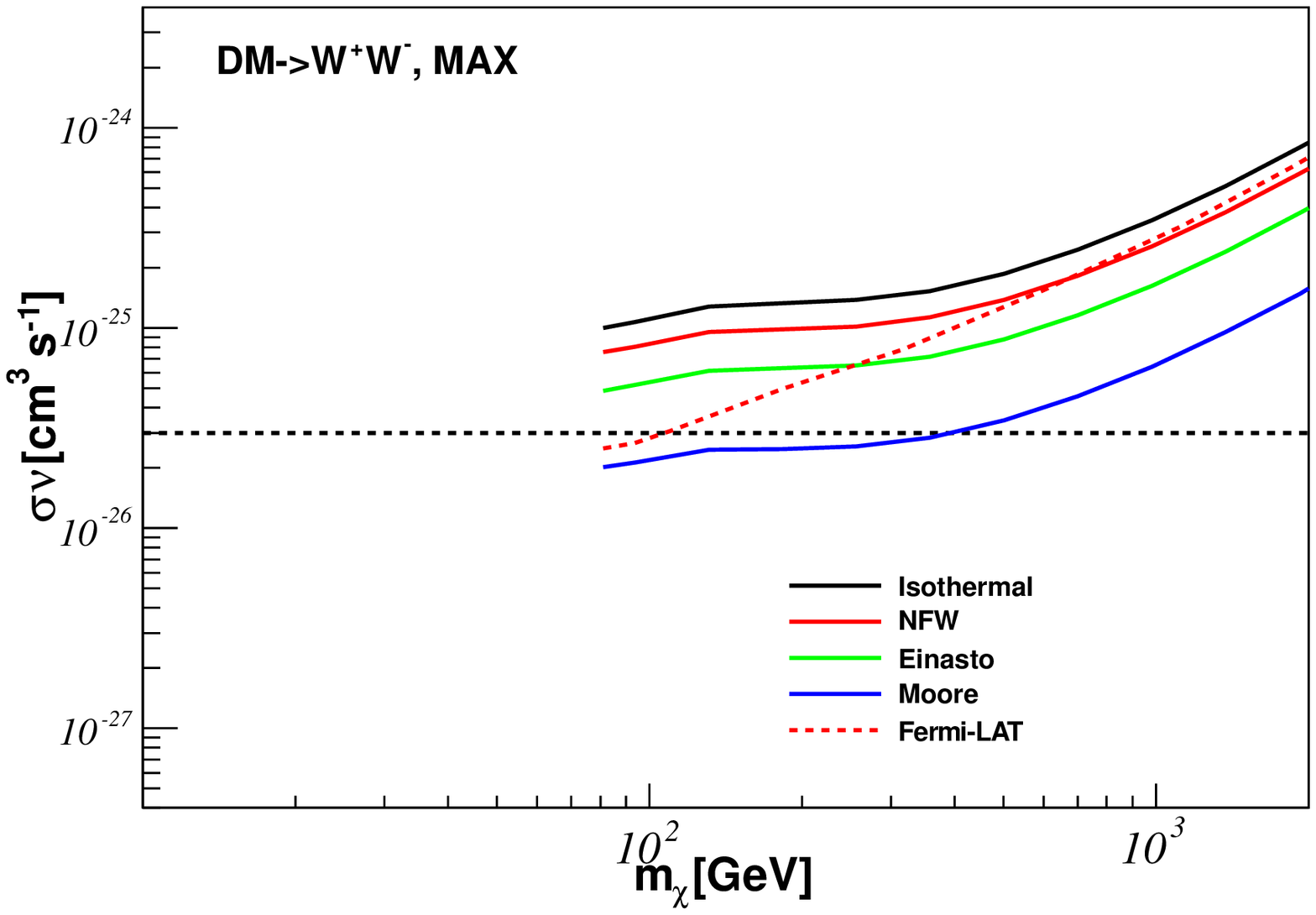}
\caption{The same as \fig{fig:bb}, but for DM annihilation into $W^{+}W^{-}$ final states.}
\label{fig:WW}
\end{center}
\end{figure}
%
In \fig{fig:bb}, 
we show the obtained upper limits on 
the cross sections for 
DM particle annihilation into $b\bar b$ final states from
the AMS-02 $\bar p/p$ data in 
the ``conventional'', ``MED'',  ``MIN''  and ``MAX'' propagation models.
Four different DM profiles: 
NFW~\cite{Navarro:1996gj}, 
Isothermal~\cite{Bergstrom:1997fj}, 
Einasto~\cite{Einasto:2009zd
}
and 
Moore~\cite{
Moore:1999nt, 
%
Diemand:2004wh} 
are considered.
%
As can be seen, 
the upper limits as a function of $m_{\chi}$ show some smooth structure for
all the final states and DM profiles.
The limits tend to be relatively stronger at $m_{\chi}\approx 300$ GeV,
which is related to the fact that the background predictions agree with 
the data well at the antiproton energy range $\sim 20-100$ GeV. 
For a comparison,
the upper limits  from 
the Fermi-LAT 6-year gamma-ray data of 
the dwarf spheroidal satellite galaxies of the Milky Way~\cite{Ackermann:2015zua}
are also shown in \fig{fig:bb}.
In the ``conventional'' model, 
the upper limits from the AMS-02 $\bar p/p$ data are found to be compatible with that derived from
the Fermi-LAT gamma-ray data for $m_{\chi}\gtrsim 300$ GeV. 
This observation holds for most of the DM profiles.
In the ``MED'' model, the constraints are relatively weaker,
which is related to the under prediction of low energy antiprotons
in this model and the limits are more conservative.
For an estimation of the uncertainties due to the propagation models,
from the ``MIN'' model to the ``MAX'' model, 
we find that the variation of the upper limits is within about a factor of five.  
%

%
\begin{figure}[tb]
\begin{center}
\includegraphics[width=0.45\textwidth]{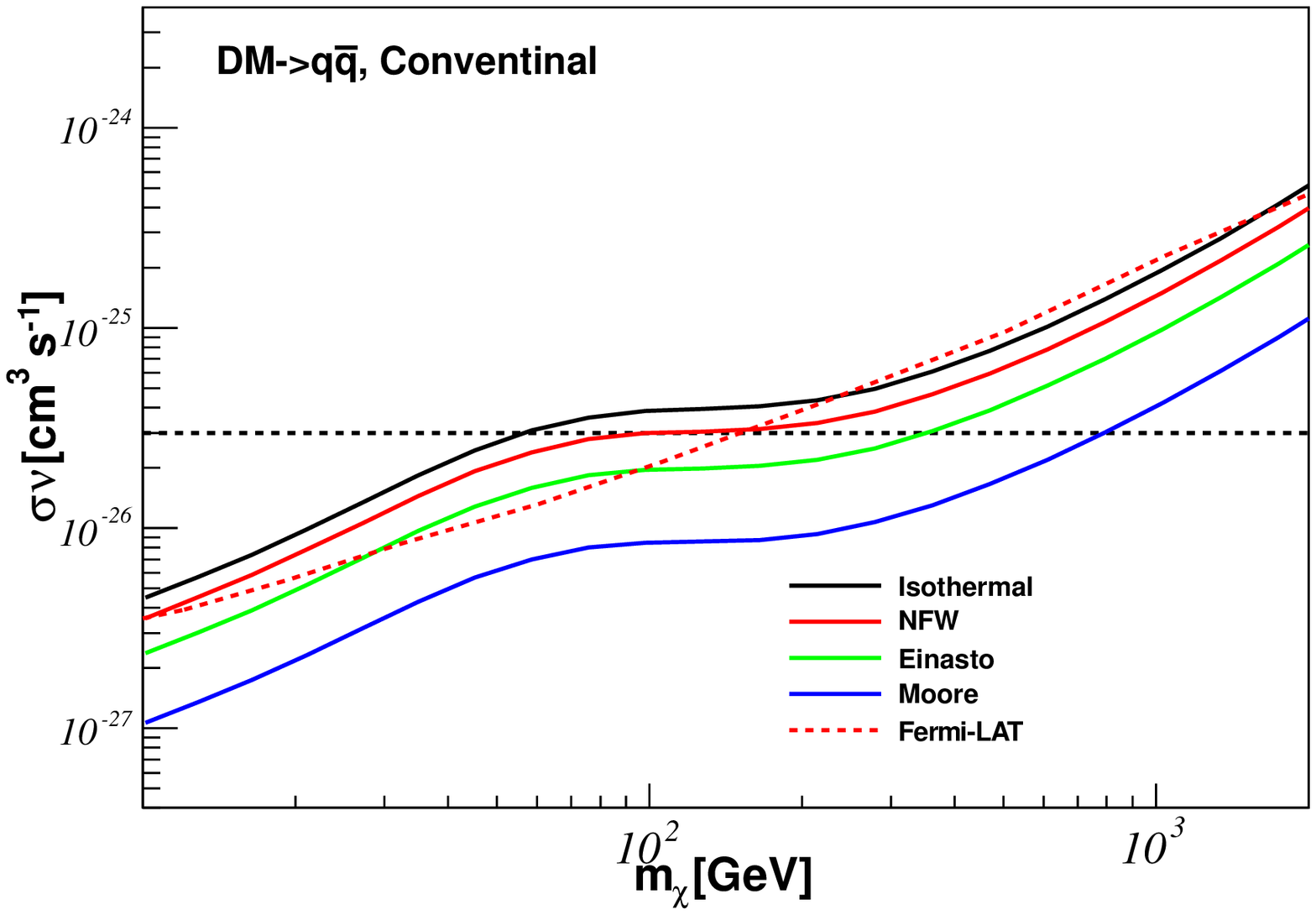}
\includegraphics[width=0.45\textwidth]{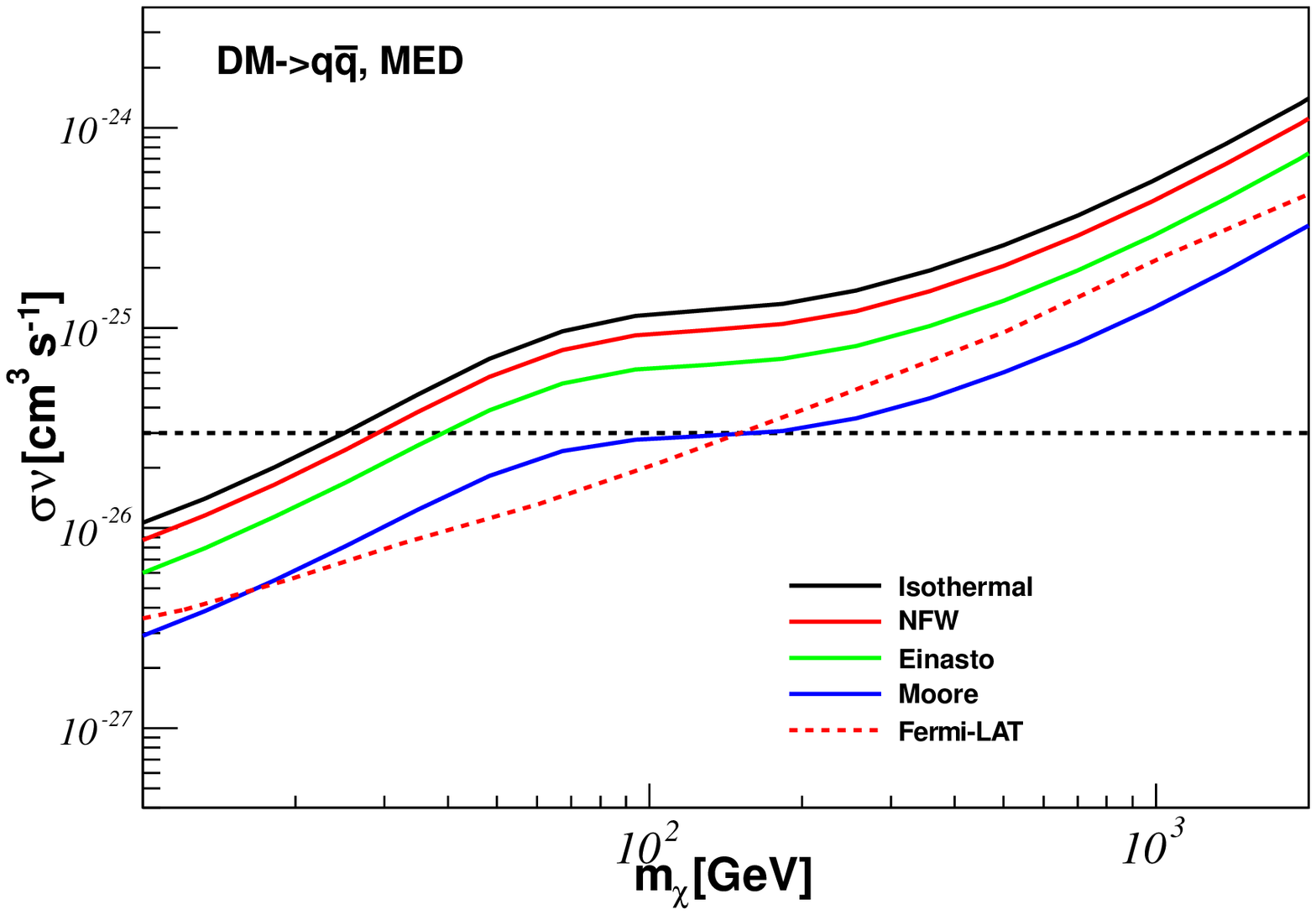}
\\
\includegraphics[width=0.45\textwidth]{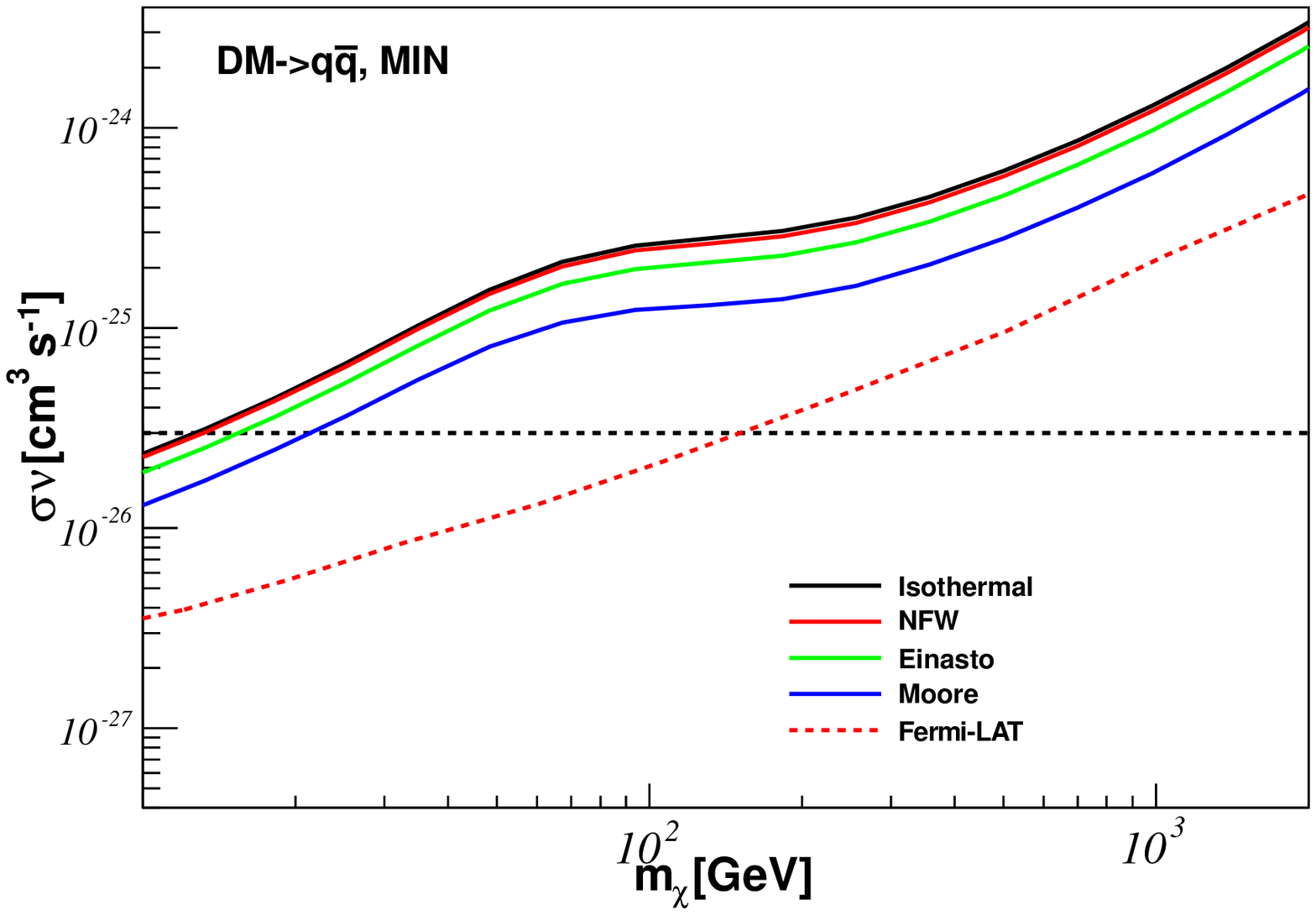}
\includegraphics[width=0.45\textwidth]{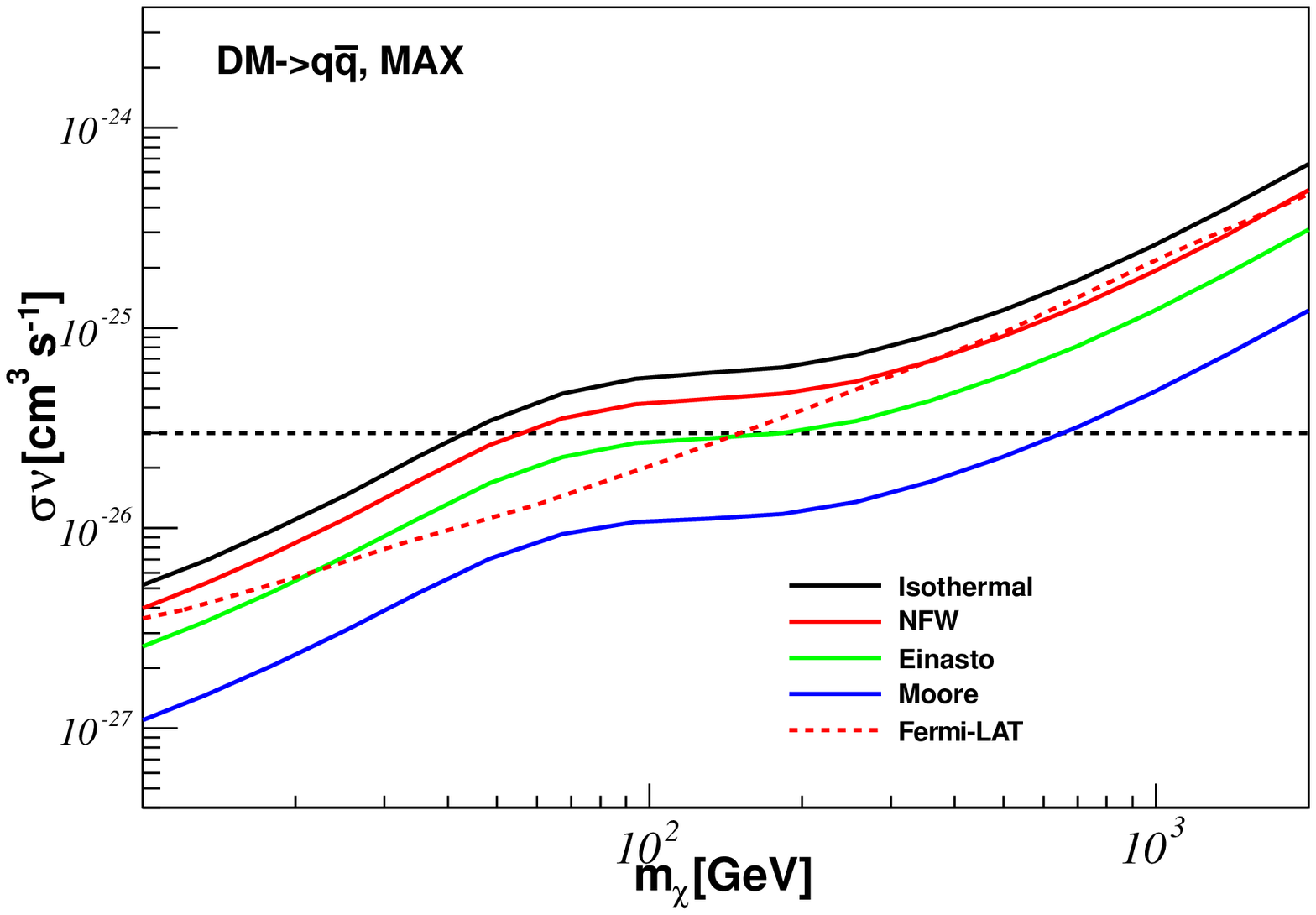}
\caption{The same as \fig{fig:bb}, but for DM annihilation into $q\bar q$ final states.}
\label{fig:qq}
\end{center}
\end{figure}
%

For the $W^{+}W^{-}$ final states, 
the resulting limits are shown in \fig{fig:WW}.
In the ``conventional'' propagation model,
the constraints from AMS-02 $\bar p/p$ data turn out to be
more stringent than that from 
the Fermi-LAT gamma-ray data for 
all the four DM profiles 
when the DM particle mass is above $\sim300$ GeV.
Again we find that the variation of the upper limits from the ``MIN'' to the ``MAX''
model is within a factor of five.
The result for the $q\bar q$ final states is shown in \fig{fig:qq}.
%
%
Compared with the case of $W^{+}W^{-}$ and $b\bar b$, 
the constraints on the  $q\bar q$ final states are the most stringent.
For all the three final states, 
we find that 
the allowed DM annihilation cross section is below 
the typical thermal cross section for $m_{\chi}\lesssim 300$ GeV in 
the conventional propagation model with Einasto profile.

%
\begin{figure}[tbh]
\begin{center}
\includegraphics[width=0.45\textwidth]{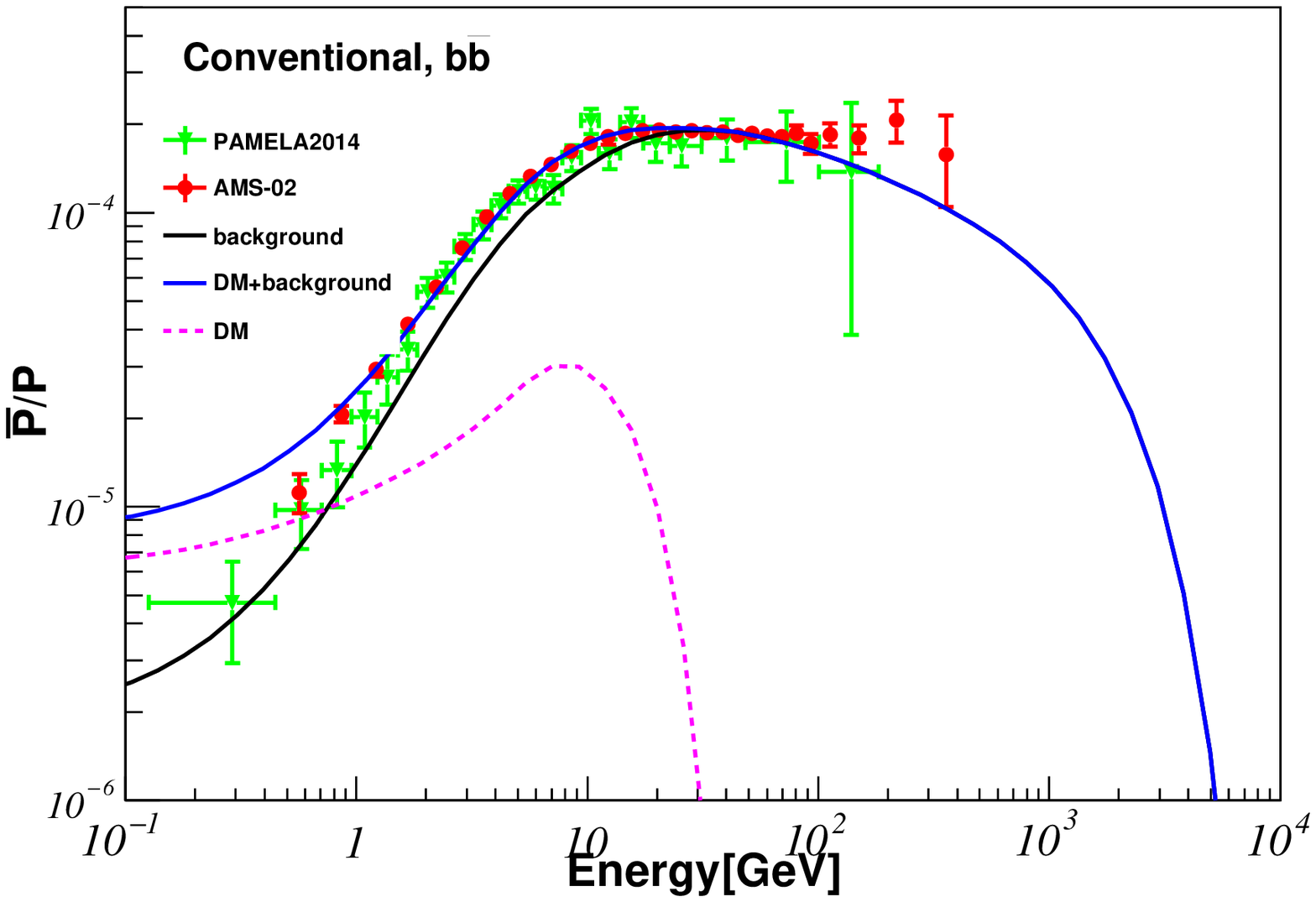}
\includegraphics[width=0.45\textwidth]{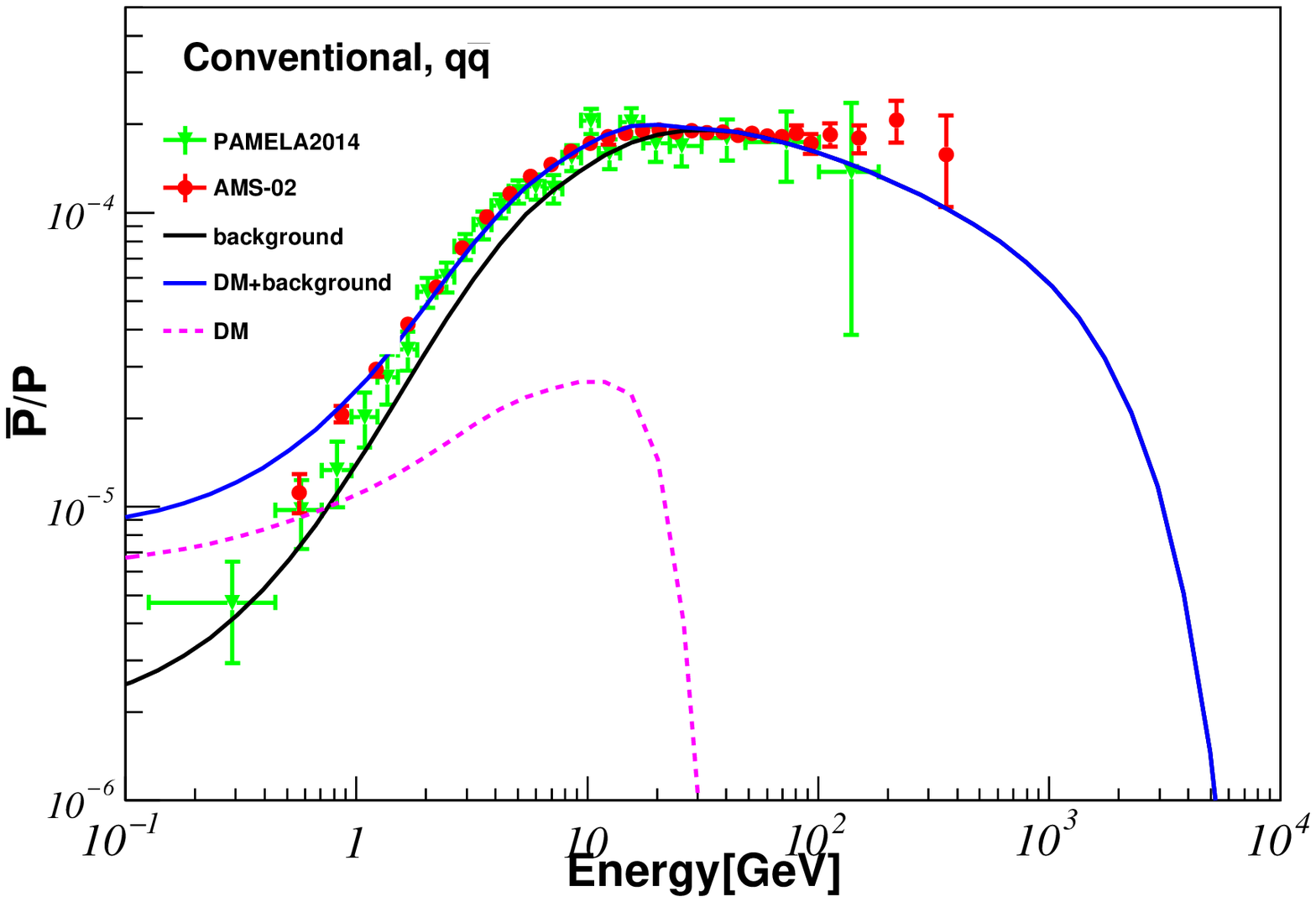}
\caption{ 
Left) 
Spectrum of  $\bar p/p$ flux ratio  from DM annihilating into  $\bar b b$ final states with
$m_{\chi}=58.5 $~GeV and 
$\langle \sigma v \rangle=2.16\times 10^{-26}\text{ cm}^{3}\text{s}^{-1}$ obtained from
a fit to the whole AMS-02  $\bar p/p$ data~\cite{Ting:AMS}. 
The ``conventional'' background model and the Einasto DM profile are assumed.
Right) The same as left, but for the fit with $\bar q q$ final state with 
the best-fit values
$m_{\chi}=35$~GeV and 
$\langle \sigma v \rangle=0.86\times 10^{-26}\text{ cm}^{3}\text{s}^{-1}$.
}
\label{fig:low-energy-fit}
\end{center}
\end{figure}
%

%
\begin{figure}[htb]
\begin{center}
\includegraphics[width=0.45\textwidth]{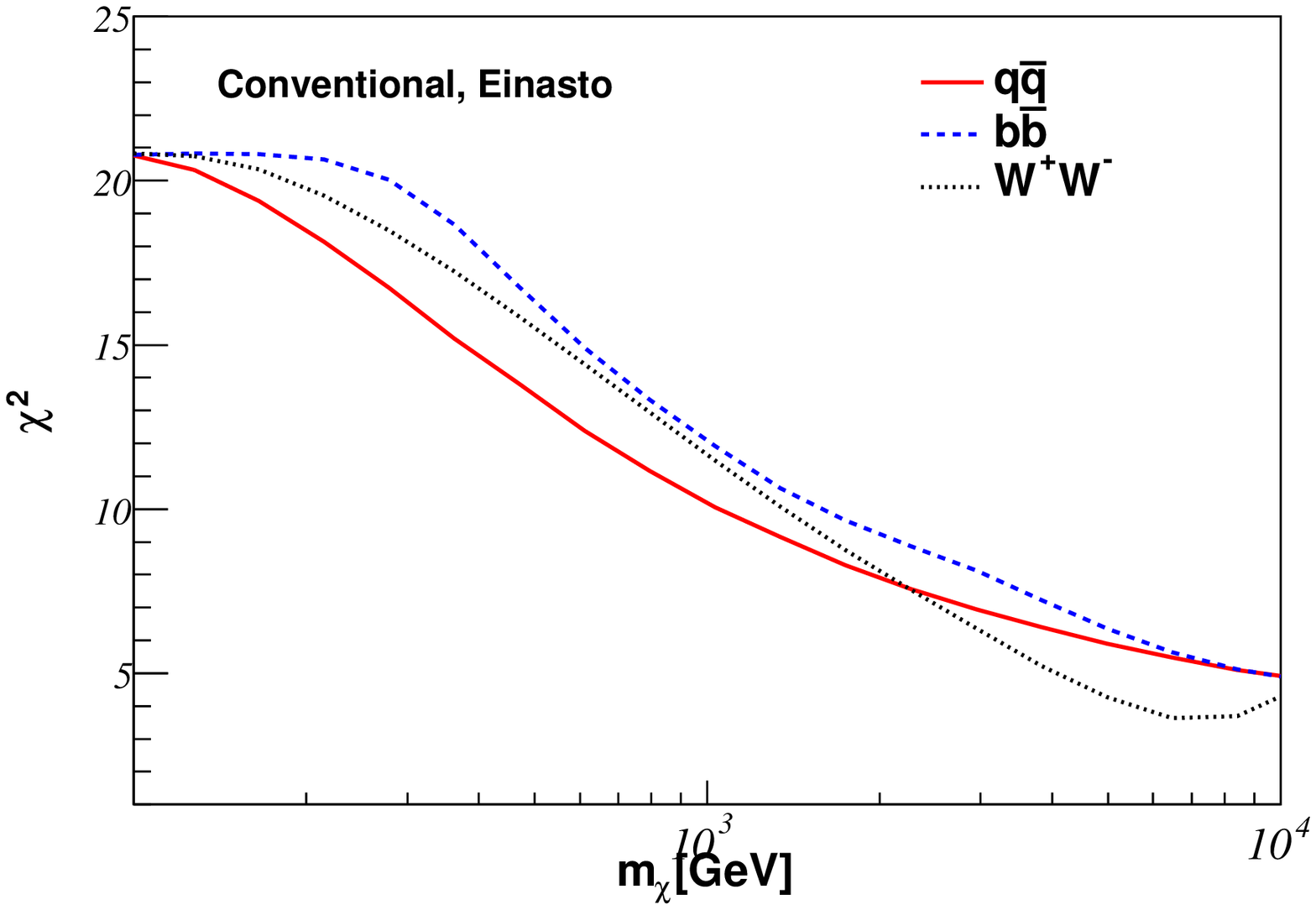}
\includegraphics[width=0.45\textwidth]{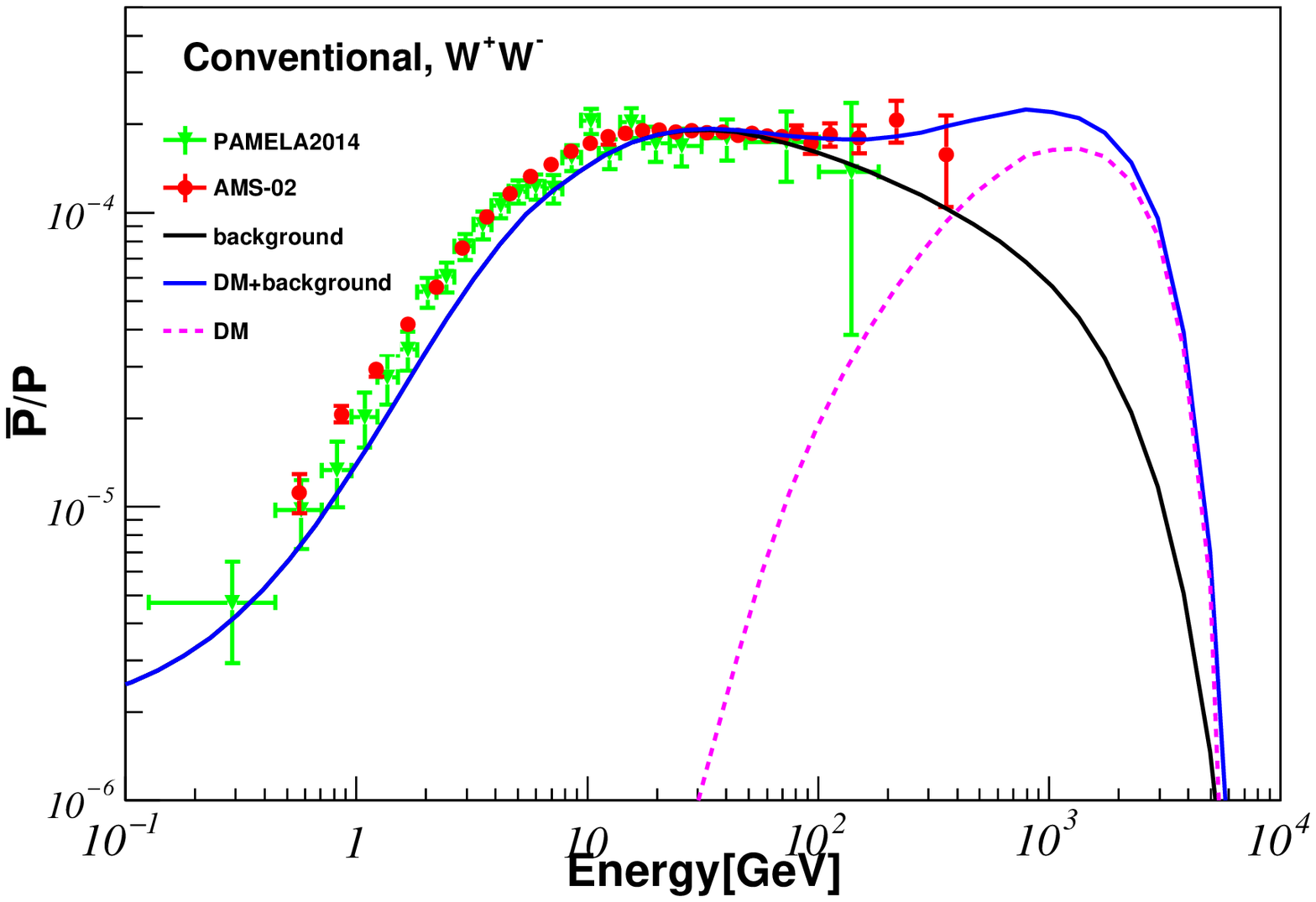}
\\
\includegraphics[width=0.45\textwidth]{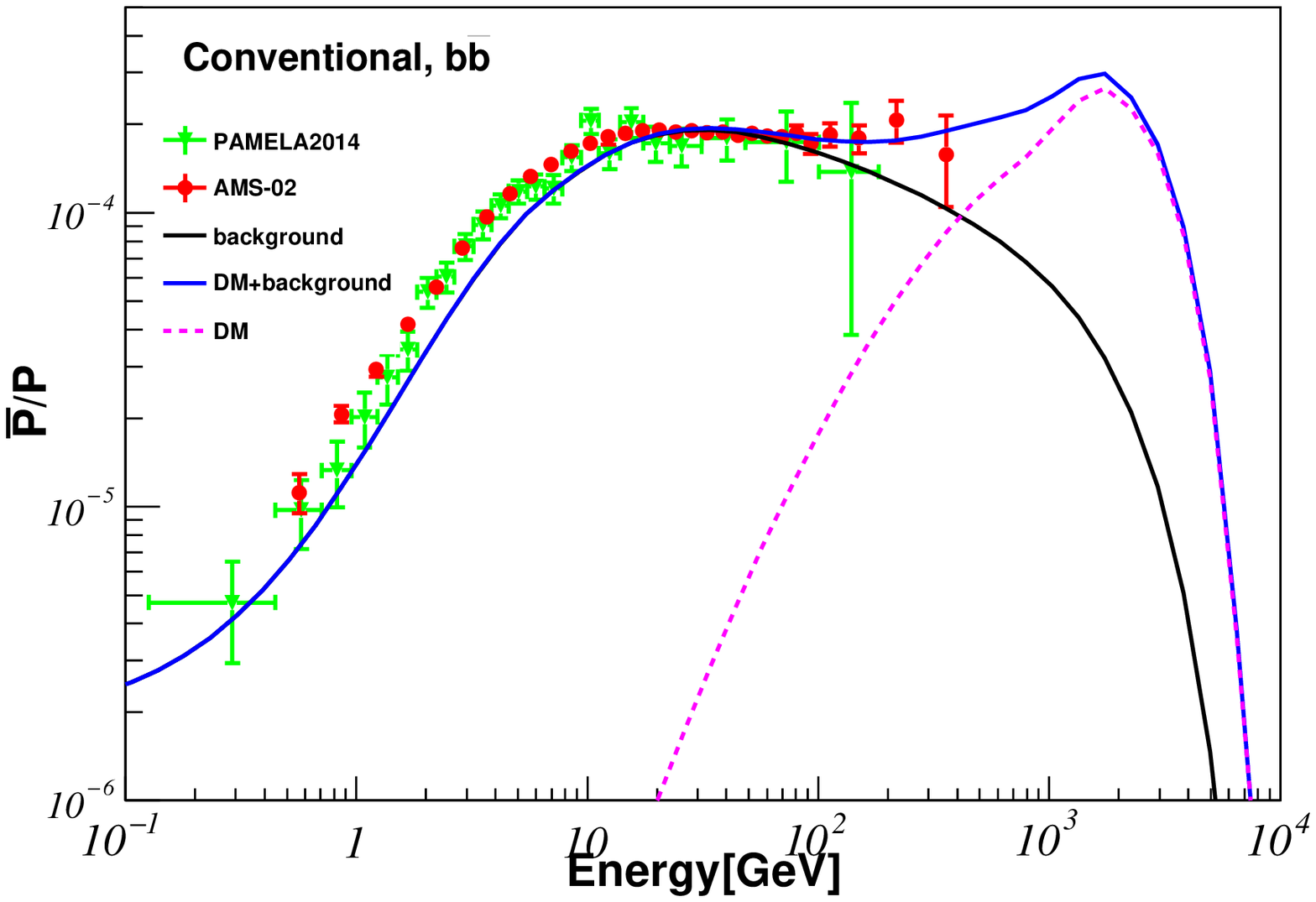}
\includegraphics[width=0.45\textwidth]{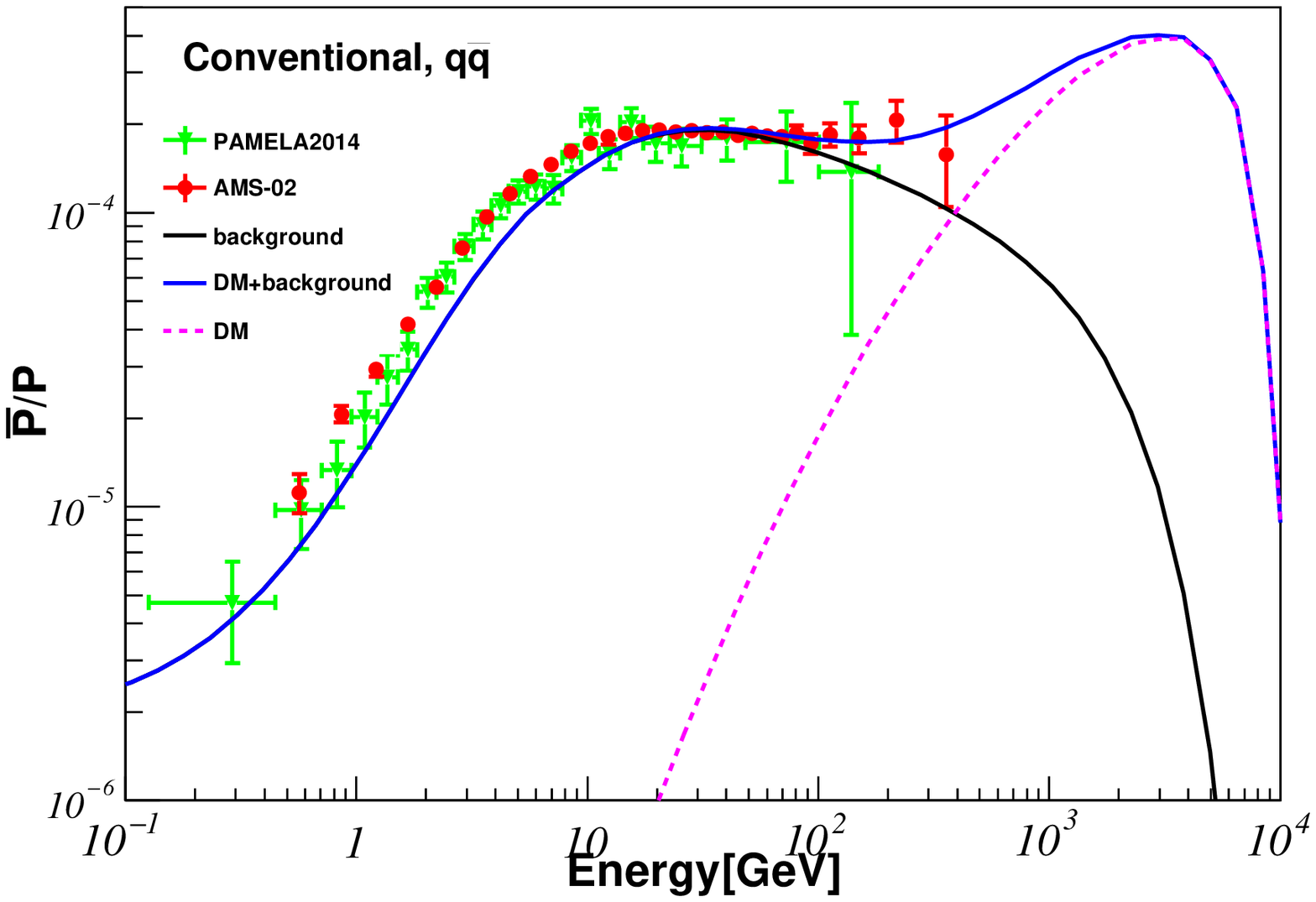}
\caption{
(Upper left)
values of $\chi^{2}_{\text{min}}$ as a function of 
DM particle mass $m_{\chi}$ from 
a fit to the AMS-02 $\bar p/p$ data ( with kinetic energy above 20 GeV ) in 
the ``conventional'' propagation model
~\cite{astro-ph/0101068,astro-ph/0510335}  with  
the DM profile  fixed to  Einasto~\cite{Einasto:2009zd
}.
Three annihilation channels  $b\bar b$, $q\bar q$ and 
$W^{+}W^{-}$ are considered.
(Upper right)
predicted $\bar p/p$ ratio in the case of 
background (``conventional'' model)
plus a DM contribution with 
$m_{\chi}=6.5$~TeV,
$\langle \sigma v\rangle=1.9\times 10^{-24}\text{cm}^{3}\text{s}^{-1}$,
and annihilation final states $W^{+}W^{-}$.
The flux ratio of antiproton from DM to the proton from the  background  
$\bar p_{\text{DM}}/p_{\text{BG}}$ is shown as the dashed line.
The data  from AMS-02~\cite{Ting:AMS}
and PAMELA~\cite{Adriani:2014pza} are also shown.
(Lower left)
the same as the upper right, but for the $b\bar b$ channel
with
$m_{\chi}=10.9$~TeV and 
$\langle \sigma v\rangle=3.4\times 10^{-24}~\text{cm}^{3}\text{s}^{-1}$.
(Lower right)
the same as the upper right, but for the $q\bar q$ channel with
$m_{\chi}=10.9$~TeV and 
$\langle \sigma v\rangle=3.3\times 10^{-24}~\text{cm}^{3}\text{s}^{-1}$.
}
\label{fig:chisq}
\end{center}
\end{figure}
%

As shown in \fig{fig:pbar}, 
compared with the AMS-02 data
the GALPROP  DR models predict fewer antiprotons at
low ($\lesssim 10$ GeV) and very high ($\gtrsim 100$ GeV ) energies.
Without a robust estimation of the theoretical uncertainties,
it is too early  to claim any excesses in the $\bar p/p$ data.
We nevertheless consider what would be the implications for DM  
if such a trend in the observations is confirmed by  future analyses. 
%
The low energy data would allow for a non-vanishing DM annihilation cross section.
For instance, 
in  the ``conventional'' propagation model, 
for $m_{\chi}=$10.1, 35.0 and 75.8~GeV, 
the best-fit  values are
$\langle \sigma v\rangle =3.6\times 10^{-27}$,
$1.14\times 10^{-26}$, and 
$2.79\times 10^{-26}\text{ cm}^{3}\text{s}^{-1}$, 
respectively, 
if the DM profile is Einasto, 
and the DM particles annihilate dominantly into $\bar b b$ final states.
If both $m_{\chi}$ and $\langle \sigma v\rangle$ are allowed to vary freely,
the best-fit  DM particle masses and annihilation cross sections  are 
$m_{\chi}=58.5\ (35.0)~\text{GeV and } 
\langle \sigma v\rangle=2.16\ (0.86)\times 10^{-26}\text{ cm}^{3}\text{s}^{-1}
$
for DM annihilating into $\bar b b\ (\bar q q)$ final states.
In \fig{fig:low-energy-fit}, 
we show the calculated spectra of $\bar p/p$ flux ratio from 
the best-fit DM particle masses and cross sections.
The figure shows that the low energy $\bar p/p$ data are well reproduced
by including such a DM contribution,
except for the data point with kinetic energy below 1~GeV.
%

As shown in \fig{fig:pbar}, 
the spectrum of the  AMS-02 $\bar p/p$ ratio tends to be flat toward 
high energies above $\sim100$ GeV.
This trend, if confirmed by the future AMS-02 data,
is not expected from the secondary production of antiprotons,
and 
raises the interesting question whether 
this would leave some room for a heavy DM contribution,
similar to the case of the AMS-02 positron fraction
\cite{Accardo:2014lma, 
Kopp:2013eka,
Bergstrom:2013jra, 
Jin:2013nta,Liu:2013vha}.
To explore this possibility,
we perform an other  fit using
the $\bar p/p$ ratio data above 20~GeV (including 15 data points in total)
in order to avoid the theoretical uncertainties in the low energy region.
The obtained  
$\chi^{2}_{\text{min}}$ as a function of $m_{\chi}$ for
the $b\bar b$, $q\bar q$ and $W^{+}W^{-}$ final states in
the  ``conventional'' propagation model with Einasto DM profile
are shown in \fig{fig:chisq}.
%
For the three final states
the values of $\chi^{2}_{\text{min}}$  decrease almost monotonically 
from $\sim21$ to $\sim5$ with 
an increasing DM particles mass from $100$~GeV to $10$~TeV,
but
the $\chi^{2}$-curves become gradually  flat toward high DM masses.
Only for the $W^{+}W^{-}$ channel, there exists a shallow  local minimal at
around 6.5~TeV with low statistical significance. 
From the $\chi^{2}$-curves, 
one can see that the DM particles mass is restricted to be
above $\sim2$~TeV at $2\sigma$. 
For an illustration purpose, 
we  show in \fig{fig:chisq} 
the predictions for the  $\bar p/p$ ratio in  
the ``conventional''  background model  with a DM contribution. 
The DM particles masses and annihilation cross sections  chosen to be 
$m_{\chi}=6.5$~TeV,
$\langle \sigma v\rangle=1.9\times 10^{-24}\text{cm}^{3}\text{s}^{-1}$
for $W^{+}W^{-}$,
$m_{\chi}=10.9$~TeV,
$\langle \sigma v\rangle=3.4\times 10^{-24}~\text{cm}^{3}\text{s}^{-1}$
for $b\bar b$ channel, 
and 
$m_{\chi}=10.9$~TeV and 
$\langle \sigma v\rangle=3.3\times 10^{-24}~\text{cm}^{3}\text{s}^{-1}$
for $q\bar q$ channel.
Note that these values are not from the best-fit values.
We conclude that 
introducing a DM contribution can  improve the 
agreement with the AMS-02 $\bar p/p$ data with kinetic energy above 100~GeV,
but the statistics is not high enough to
determine  the DM properties such as its mass and interaction strength.
%
%
As can be see in \fig{fig:pbar}, 
the possible ``excess'' is located at the kinetic energy range $100-450$~GeV where
the secondary backgrounds from the four propagation models are similar.
However, beyond $\sim 450$ GeV, the  $\bar p/p$ from the ``conventional'' model 
drops quicker than that in the other propagation models. 
The future high energy antiproton data will be very important not only in 
probing DM but also in constraining the background models.

In conclusion,
%
we have explored the implications of the first  AMS-02 $\bar p/p$ data on 
constraining the annihilation cross sections of the DM particles in 
various propagation models and DM profiles. 
%
%
We have  derived the upper limits using the GALPROP code
and 
shown that 
in the ``conventional '' propagation model with  Einasto DM profile, 
the constraints can be more stringent than that derived from
the Ferm-LAT gamma-ray data on 
the dwarf spheroidal satellite galaxies.
Making use of  the typical 
minimal, median and maximal models obtained from 
a previous global fit, we have shown that
the uncertainties on the upper limits is around a factor of five.
The future more precise AMS-02 data can help to reduce the uncertainties 
in the derived upper limits. 
%
%
%
The analysis in this work has some overlap with that in \cite{Giesen:2015ufa} .
Note that although the conclusions are similar, 
the analysis in this  work  is based on the fully numerical GALPROP code, 
while that in \cite{Giesen:2015ufa}  is based on the two-zone diffusion model
with (semi)-analytical approach. 
%
%
%
Similar discussions on the DM matter contributions can be found in 
Refs.~\cite{
Ibe:2015tma,
Chen:2015cqa%
}.

%
This work is supported in part by
the National Basic Research Program of China (973 Program) under Grants 
No. 2010CB833000;
the National Nature Science Foundation of China (NSFC) under Grants 
%
%
No. 10905084,
%
No. 11335012 and
%
No. 11475237;
%
The numerical calculations were done  using
the HPC Cluster of SKLTP/ITP-CAS.

\bibliography{amsfit_inspire,misc}

\providecommand{\href}[2]{#2}\begingroup\raggedright\begin{thebibliography}{10}

\bibitem{Adriani:2008zr}
{\bf PAMELA} Collaboration, O.~Adriani et~al., {\it {An anomalous positron
  abundance in cosmic rays with energies 1.5-100 GeV}},  {\em Nature} {\bf 458}
  (2009) 607--609, [\href{http://arxiv.org/abs/0810.4995}{{\tt
  arXiv:0810.4995}}].

\bibitem{Adriani:2010ib}
O.~Adriani, G.~Barbarino, G.~Bazilevskaya, R.~Bellotti, M.~Boezio, et~al., {\it
  {A statistical procedure for the identification of positrons in the PAMELA
  experiment}},  {\em Astropart.Phys.} {\bf 34} (2010) 1--11,
  [\href{http://arxiv.org/abs/1001.3522}{{\tt arXiv:1001.3522}}].

\bibitem{FermiLAT:2011ab}
{\bf Fermi-LAT} Collaboration, M.~Ackermann et~al., {\it {Measurement of
  separate cosmic-ray electron and positron spectra with the Fermi Large Area
  Telescope}},  {\em Phys.Rev.Lett.} {\bf 108} (2012) 011103,
  [\href{http://arxiv.org/abs/1109.0521}{{\tt arXiv:1109.0521}}].

\bibitem{Accardo:2014lma}
{\bf AMS} Collaboration, L.~Accardo et~al., {\it {High Statistics Measurement
  of the Positron Fraction in Primary Cosmic Rays of 0.5¨C500 GeV with the
  Alpha Magnetic Spectrometer on the International Space Station}},  {\em
  Phys.Rev.Lett.} {\bf 113} (2014) 121101.

\bibitem{Kopp:2013eka}
J.~Kopp, {\it {Constraints on dark matter annihilation from AMS-02 results}},
  {\em Phys.Rev.} {\bf D88} (2013) 076013,
  [\href{http://arxiv.org/abs/1304.1184}{{\tt arXiv:1304.1184}}].

\bibitem{DeSimone:2013fia}
A.~De~Simone, A.~Riotto, and W.~Xue, {\it {Interpretation of AMS-02 Results:
  Correlations among Dark Matter Signals}},  {\em JCAP} {\bf 1305} (2013) 003,
  [\href{http://arxiv.org/abs/1304.1336}{{\tt arXiv:1304.1336}}].

\bibitem{Cholis:2013psa}
I.~Cholis and D.~Hooper, {\it {Dark Matter and Pulsar Origins of the Rising
  Cosmic Ray Positron Fraction in Light of New Data From AMS}},  {\em
  Phys.Rev.} {\bf D88} (2013) 023013,
  [\href{http://arxiv.org/abs/1304.1840}{{\tt arXiv:1304.1840}}].

\bibitem{Feng:2013zca}
L.~Feng, R.-Z. Yang, H.-N. He, T.-K. Dong, Y.-Z. Fan, et~al., {\it {AMS-02
  positron excess: new bounds on dark matter models and hint for primary
  electron spectrum hardening}},  {\em Phys.Lett.} {\bf B728} (2014) 250--255,
  [\href{http://arxiv.org/abs/1303.0530}{{\tt arXiv:1303.0530}}].

\bibitem{Jin:2013nta}
H.-B. Jin, Y.-L. Wu, and Y.-F. Zhou, {\it {Implications of the first AMS-02
  measurement for dark matter annihilation and decay}},  {\em JCAP} {\bf 1311}
  (2013) 026, [\href{http://arxiv.org/abs/1304.1997}{{\tt arXiv:1304.1997}}].

\bibitem{Liu:2013vha}
Z.-P. Liu, Y.-L. Wu, and Y.-F. Zhou, {\it {Sommerfeld enhancements with vector,
  scalar and pseudoscalar force-carriers}},  {\em Phys.Rev.} {\bf D88} (2013)
  096008, [\href{http://arxiv.org/abs/1305.5438}{{\tt arXiv:1305.5438}}].

\bibitem{Bergstrom:2013jra}
L.~Bergstrom, T.~Bringmann, I.~Cholis, D.~Hooper, and C.~Weniger, {\it {New
  limits on dark matter annihilation from AMS cosmic ray positron data}},  {\em
  Phys.Rev.Lett.} {\bf 111} (2013) 171101,
  [\href{http://arxiv.org/abs/1306.3983}{{\tt arXiv:1306.3983}}].

\bibitem{Ibarra:2013zia}
A.~Ibarra, A.~S. Lamperstorfer, and J.~Silk, {\it {Dark matter annihilations
  and decays after the AMS-02 positron measurements}},  {\em Phys.Rev.} {\bf
  D89} (2014), no.~6 063539, [\href{http://arxiv.org/abs/1309.2570}{{\tt
  arXiv:1309.2570}}].

\bibitem{DiMauro:2014iia}
M.~Di~Mauro, F.~Donato, N.~Fornengo, R.~Lineros, and A.~Vittino, {\it
  {Interpretation of AMS-02 electrons and positrons data}},  {\em JCAP} {\bf
  1404} (2014) 006, [\href{http://arxiv.org/abs/1402.0321}{{\tt
  arXiv:1402.0321}}].

\bibitem{Lin:2014vja}
S.-J. Lin, Q.~Yuan, and X.-J. Bi, {\it {Quantitative study of the AMS-02
  electron/positron spectra: Implications for pulsars and dark matter
  properties}},  {\em Phys.Rev.} {\bf D91} (2015), no.~6 063508,
  [\href{http://arxiv.org/abs/1409.6248}{{\tt arXiv:1409.6248}}].

\bibitem{Ibe:2014qya}
M.~Ibe, S.~Matsumoto, S.~Shirai, and T.~T. Yanagida, {\it {Mass of Decaying
  Wino from AMS-02 2014}},  {\em Phys.Lett.} {\bf B741} (2015) 134--137,
  [\href{http://arxiv.org/abs/1409.6920}{{\tt arXiv:1409.6920}}].

\bibitem{Ting:AMS}
S.Ting, talk at {\it AMS-02 days at CERN}, April 15-17, CERN, Geneva,
  https://indico.cern.ch/event/381134/timetable/\#20150415.

\bibitem{Hooper:2014ysa}
D.~Hooper, T.~Linden, and P.~Mertsch, {\it {What Does The PAMELA Antiproton
  Spectrum Tell Us About Dark Matter?}},  {\em JCAP} {\bf 1503} (2015), no.~03
  021, [\href{http://arxiv.org/abs/1410.1527}{{\tt arXiv:1410.1527}}].

\bibitem{Kappl:2014hha}
R.~Kappl and M.~W. Winkler, {\it {The Cosmic Ray Antiproton Background for
  AMS-02}},  {\em JCAP} {\bf 1409} (2014) 051,
  [\href{http://arxiv.org/abs/1408.0299}{{\tt arXiv:1408.0299}}].

\bibitem{Fornengo:2013xda}
N.~Fornengo, L.~Maccione, and A.~Vittino, {\it {Constraints on particle dark
  matter from cosmic-ray antiprotons}},  {\em JCAP} {\bf 1404} (2014) 003,
  [\href{http://arxiv.org/abs/1312.3579}{{\tt arXiv:1312.3579}}].

\bibitem{Cirelli:2013hv}
M.~Cirelli and G.~Giesen, {\it {Antiprotons from Dark Matter: Current
  constraints and future sensitivities}},  {\em JCAP} {\bf 1304} (2013) 015,
  [\href{http://arxiv.org/abs/1301.7079}{{\tt arXiv:1301.7079}}].

\bibitem{Jin:2012jn}
H.-B. Jin, S.~Miao, and Y.-F. Zhou, {\it {Implications of the latest XENON100
  and cosmic ray antiproton data for isospin violating dark matter}},  {\em
  Phys.Rev.} {\bf D87} (2013), no.~1 016012,
  [\href{http://arxiv.org/abs/1207.4408}{{\tt arXiv:1207.4408}}].

\bibitem{Jin:2015sqa}
H.-B. Jin, Y.-L. Wu, and Y.-F. Zhou, {\it {Upper limits on DM annihilation
  cross sections from the first AMS-02 antiproton data}},
  \href{http://arxiv.org/abs/1504.04604}{{\tt arXiv:1504.04604}}.

\bibitem{Ginzburg:1990sk}
V.~Ginzburg, V.~Dogiel, V.~Berezinsky, S.~Bulanov, and V.~Ptuskin, {\it
  {Astrophysics of cosmic rays}}, .

\bibitem{Strong:1998pw}
A.~Strong and I.~Moskalenko, {\it {Propagation of cosmic-ray nucleons in the
  galaxy}},  {\em Astrophys.J.} {\bf 509} (1998) 212--228,
  [\href{http://arxiv.org/abs/astro-ph/9807150}{{\tt astro-ph/9807150}}].

\bibitem{Gleeson:1968zza}
L.~Gleeson and W.~Axford, {\it {Solar Modulation of Galactic Cosmic Rays}},
  {\em Astrophys.J.} {\bf 154} (1968) 1011.

\bibitem{astro-ph/9807150}
A.~Strong and I.~Moskalenko, {\it {Propagation of cosmic-ray nucleons in the
  galaxy}},  {\em Astrophys.J.} {\bf 509} (1998) 212--228,
  [\href{http://arxiv.org/abs/astro-ph/9807150}{{\tt astro-ph/9807150}}].

\bibitem{astro-ph/0106567}
I.~V. Moskalenko, A.~W. Strong, J.~F. Ormes, and M.~S. Potgieter, {\it
  {Secondary anti-protons and propagation of cosmic rays in the galaxy and
  heliosphere}},  {\em Astrophys.J.} {\bf 565} (2002) 280--296,
  [\href{http://arxiv.org/abs/astro-ph/0106567}{{\tt astro-ph/0106567}}].

\bibitem{astro-ph/0101068}
A.~Strong and I.~Moskalenko, {\it {Models for galactic cosmic ray
  propagation}},  {\em Adv.Space Res.} {\bf 27} (2001) 717--726,
  [\href{http://arxiv.org/abs/astro-ph/0101068}{{\tt astro-ph/0101068}}].

\bibitem{astro-ph/0210480}
I.~V. Moskalenko, A.~Strong, S.~Mashnik, and J.~Ormes, {\it {Challenging cosmic
  ray propagation with antiprotons. Evidence for a fresh nuclei component?}},
  {\em Astrophys.J.} {\bf 586} (2003) 1050--1066,
  [\href{http://arxiv.org/abs/astro-ph/0210480}{{\tt astro-ph/0210480}}].

\bibitem{astro-ph/0510335}
V.~Ptuskin, I.~V. Moskalenko, F.~Jones, A.~Strong, and V.~Zirakashvili, {\it
  {Dissipation of magnetohydrodynamic waves on energetic particles: impact on
  interstellar turbulence and cosmic ray transport}},  {\em Astrophys.J.} {\bf
  642} (2006) 902--916, [\href{http://arxiv.org/abs/astro-ph/0510335}{{\tt
  astro-ph/0510335}}].

\bibitem{Jin:2014ica}
H.-B. Jin, Y.-L. Wu, and Y.-F. Zhou, {\it {Cosmic ray propagation and dark
  matter in light of the latest AMS-02 data}},
  \href{http://arxiv.org/abs/1410.0171}{{\tt arXiv:1410.0171}}.

\bibitem{Donato:2003xg}
F.~Donato, N.~Fornengo, D.~Maurin, and P.~Salati, {\it {Antiprotons in cosmic
  rays from neutralino annihilation}},  {\em Phys.Rev.} {\bf D69} (2004)
  063501, [\href{http://arxiv.org/abs/astro-ph/0306207}{{\tt
  astro-ph/0306207}}].

\bibitem{Moskalenko:2001ya}
I.~V. Moskalenko, A.~W. Strong, J.~F. Ormes, and M.~S. Potgieter, {\it
  {Secondary anti-protons and propagation of cosmic rays in the galaxy and
  heliosphere}},  {\em Astrophys.J.} {\bf 565} (2002) 280--296,
  [\href{http://arxiv.org/abs/astro-ph/0106567}{{\tt astro-ph/0106567}}].

\bibitem{Moskalenko:2002yx}
I.~V. Moskalenko, A.~Strong, S.~Mashnik, and J.~Ormes, {\it {Challenging cosmic
  ray propagation with antiprotons. Evidence for a fresh nuclei component?}},
  {\em Astrophys.J.} {\bf 586} (2003) 1050--1066,
  [\href{http://arxiv.org/abs/astro-ph/0210480}{{\tt astro-ph/0210480}}].

\bibitem{Giesen:2015ufa}
G.~Giesen, M.~Boudaud, Y.~Genolini, V.~Poulin, M.~Cirelli, et~al., {\it {AMS-02
  antiprotons, at last! Secondary astrophysical component and immediate
  implications for Dark Matter}},  \href{http://arxiv.org/abs/1504.04276}{{\tt
  arXiv:1504.04276}}.

\bibitem{Adriani:2014pza}
O.~Adriani, G.~Barbarino, G.~Bazilevskaya, R.~Bellotti, M.~Boezio, et~al., {\it
  {The PAMELA Mission: Heralding a new era in precision cosmic ray physics}},
  {\em Phys.Rept.} {\bf 544} (2014) 323--370.

\bibitem{Sjostrand:2007gs}
T.~Sjostrand, S.~Mrenna, and P.~Z. Skands, {\it {A Brief Introduction to PYTHIA
  8.1}},  {\em Comput.Phys.Commun.} {\bf 178} (2008) 852--867,
  [\href{http://arxiv.org/abs/0710.3820}{{\tt arXiv:0710.3820}}].

\bibitem{Navarro:1996gj}
J.~F. Navarro, C.~S. Frenk, and S.~D. White, {\it {A Universal density profile
  from hierarchical clustering}},  {\em Astrophys.J.} {\bf 490} (1997)
  493--508, [\href{http://arxiv.org/abs/astro-ph/9611107}{{\tt
  astro-ph/9611107}}].

\bibitem{Bergstrom:1997fj}
L.~Bergstrom, P.~Ullio, and J.~H. Buckley, {\it {Observability of gamma-rays
  from dark matter neutralino annihilations in the Milky Way halo}},  {\em
  Astropart.Phys.} {\bf 9} (1998) 137--162,
  [\href{http://arxiv.org/abs/astro-ph/9712318}{{\tt astro-ph/9712318}}].

\bibitem{Einasto:2009zd}
J.~Einasto, {\it {Dark Matter}},  \href{http://arxiv.org/abs/0901.0632}{{\tt
  arXiv:0901.0632}}.

\bibitem{Moore:1999nt}
B.~Moore, S.~Ghigna, F.~Governato, G.~Lake, T.~R. Quinn, et~al., {\it {Dark
  matter substructure within galactic halos}},  {\em Astrophys.J.} {\bf 524}
  (1999) L19--L22, [\href{http://arxiv.org/abs/astro-ph/9907411}{{\tt
  astro-ph/9907411}}].

\bibitem{Diemand:2004wh}
J.~Diemand, B.~Moore, and J.~Stadel, {\it {Convergence and scatter of cluster
  density profiles}},  {\em Mon.Not.Roy.Astron.Soc.} {\bf 353} (2004) 624,
  [\href{http://arxiv.org/abs/astro-ph/0402267}{{\tt astro-ph/0402267}}].

\bibitem{Ackermann:2015zua}
{\bf Fermi-LAT} Collaboration, M.~Ackermann et~al., {\it {Searching for Dark
  Matter Annihilation from Milky Way Dwarf Spheroidal Galaxies with Six Years
  of Fermi-LAT Data}},  \href{http://arxiv.org/abs/1503.02641}{{\tt
  arXiv:1503.02641}}.

\bibitem{Ibe:2015tma}
M.~Ibe, S.~Matsumoto, S.~Shirai, and T.~T. Yanagida, {\it {Wino Dark Matter in
  light of the AMS-02 2015 Data}},  {\em Phys. Rev.} {\bf D91} (2015), no.~11
  111701, [\href{http://arxiv.org/abs/1504.05554}{{\tt arXiv:1504.05554}}].

\bibitem{Chen:2015cqa}
C.-H. Chen, C.-W. Chiang, and T.~Nomura, {\it {Dark matter for excess of AMS-02
  positrons and antiprotons}},  {\em Phys. Lett.} {\bf B747} (2015) 495--499,
  [\href{http://arxiv.org/abs/1504.07848}{{\tt arXiv:1504.07848}}].

\end{thebibliography}\endgroup
\bibliographystyle{JHEP}

\end{document}